\documentclass{aa}     

\usepackage{graphicx}
\usepackage{txfonts}
\usepackage{natbib}
\bibpunct{(}{)}{;}{a}{}{,}

\newcommand{\Mpc}{$h^{-1}$\thinspace Mpc}

\def\apj{ApJ}
\def\apjl{ApJL}
\def\apjs{ApJS}

\def\mnras{MNRAS}

\newcommand{\be}{\begin{equation}}
\newcommand{\ee}{\end{equation}}

\begin{document}

\title{The Sloan Great Wall. Rich clusters.} 


\author {M.  Einasto\inst{1} \and E.  Tago\inst{1} \and E.  Saar\inst{1}  
\and P. Nurmi\inst{2} \and I. Enkvist\inst{1}\and P.  Einasto\inst{1} \and 
P. Hein\"am\"aki\inst{2,3} \and L.J. Liivam\"agi\inst{1} \and E.  Tempel\inst{1,4} 
\and J.  Einasto\inst{1}  \and V.J. Mart\'{\i}nez\inst{5}
\and J. Vennik\inst{1} \and P. Pihajoki\inst{2}
}

\institute{Tartu Observatory, 61602 T\~oravere, Estonia
\and 
Tuorla Observatory, University of Turku, V\"ais\"al\"antie 20, Piikki\"o, Finland
\and 
Finnish Centre of Astronomy with ESO (FINCA), University of Turku, 
V\"ais\"al\"antie 20, Piikki\"o, Finland
\and 
Institute of Physics, Tartu University, T\"ahe 4, 51010 Tartu, Estonia
\and
Observatori Astron\`omic, Universitat de Val\`encia, Apartat
de Correus 22085, E-46071 Val\`encia, Spain 
}

\authorrunning{M. Einasto et al. }

\offprints{M. Einasto}

\date{ Received   2010 / Accepted...   }

\titlerunning{The Sloan Great Wall.}

\abstract{}
{
We present the results of the
study of the substructure and galaxy content of ten rich clusters of galaxies 
in three different superclusters of the Sloan Great Wall, the richest nearby system of 
galaxies (hereafter SGW).
}
{
We determine the substructure in clusters using the 'Mclust' package from
the 'R' statistical environment and analyse their galaxy 
content with information about colours and morphological types of galaxies. We 
analyse the distribution of the peculiar velocities of galaxies in clusters and 
calculate the peculiar velocity of the first ranked galaxy.  }
{
We show that five clusters in our sample have more than one component; in some 
clusters the different components also have different galaxy content. In other clusters 
there are distinct components in the distribution of the peculiar velocities of 
galaxies. We find that in some clusters with substructure 
the peculiar velocities of 
the first ranked galaxies are high. All clusters in our sample host  luminous 
red galaxies; in eight clusters their number exceeds ten. Luminous red 
galaxies can be found both in the central areas of clusters and in the 
outskirts, some of them have high peculiar velocities. About 1/3 of the red 
galaxies in clusters are spirals. 
The scatter of colours of red ellipticals is in most clusters
larger than that of red spirals. The fraction of red galaxies in rich clusters 
in the cores of the richest superclusters is larger than the 
fraction of red galaxies in other very rich clusters in the SGW. 
}
{
The presence of substructure in rich clusters, signs of possible mergers and
infall, and the high peculiar velocities of the first ranked 
galaxies  suggest that the clusters in our sample are not yet virialized. 
We present merger trees of dark matter haloes in an N-body simulation to 
demonstrate the formation of present-day dark matter haloes via multiple mergers 
during their evolution. In simulated dark matter haloes we find a substructure
similar to  that in observed clusters.
}


\maketitle

\section{Introduction} 
\label{sect:intro} 
Most galaxies in the Universe are located in groups and clusters of galaxies, 
which in turn reside in larger systems -- in superclusters of galaxies or 
in filaments crossing underdense regions between superclusters.  According to 
the Cold Dark Matter model, groups and clusters of galaxies form hierarchically 
through the merging of smaller systems \citep[][and references therein]{loeb2008,
2000A&A...354..761K}. The timescale of the evolution of 
groups depends on their global environment \citep{tempel09}. As a result, the 
properties of groups depend on the environment where they are embedded, richer 
and more luminous groups are located in a higher-density environment than poor, 
less luminous groups \citep{e2003a, e2003b,2005A&A...436...17E,berlind06}. 
An understanding of the properties and evolutionary state of  
groups and clusters  of galaxies in different environments, and of the 
properties of galaxies in them is important for the study of groups and 
clusters, as well as for the study of the properties and evolution of galaxies 
and  larger structures -- superclusters  of galaxies.

Detailed knowledge of properties of clusters of galaxies is also needed for 
comparison of observations with N-body calculations of the formation and 
evolution of cosmic structures. 

Galaxies and galaxy systems form because of initial density perturbations of 
different scales. Perturbations of a scale of about 100~\Mpc\footnote{$h$ is the 
Hubble constant in units of 100~km~s$^{-1}$~Mpc$^{-1}$.} give rise to the 
largest superclusters. The largest and richest superclusters, which may contain 
several tens of rich (Abell) clusters, are the largest coherent systems in the 
Universe with characteristic dimensions of up to 100~\Mpc. At large scales 
a dynamical evolution takes place at a slower rate, and the richest superclusters 
have retained the memory of the initial conditions of their formation, and of the 
early evolution of structure \citep{1987Natur.326...48K}. Rich superclusters have 
high density cores  that are absent in poor superclusters \citep{e07}. The core regions 
of the richest superclusters may contain merging X-ray clusters 
\citep{rose02,2000MNRAS.312..540B,belsole04}. The formation of rich 
superclusters had to begin earlier than that of the 
smaller structures; they are the sites of 
early star and galaxy formation \citep[e.g.][]{2005ApJ...635..832M}, and the 
first places where systems of galaxies form 
\citep[e.g.][]{2004A&A...424L..17V,2005ApJ...620L...1O}. 

Among the richest galaxy systems, the Sloan Great Wall, which is the 
richest system of galaxies in the nearby Universe 
\citep{vogeley04,gott05,nichol06}, deserves special attention. The SGW 
consists of several rich and poor superclusters connected by lower density 
filaments of galaxies, representing a variety of global environments from the high 
density core of the richest supercluster in the SGW to a lower density poor 
supercluster at the edge of the SGW. The superclusters in the SGW differ in 
morphology and galaxy content, which suggests that their formation and evolution has 
been different (Einasto et al. 2010, in preparation, hereafter E10). 

Our aim in the present paper  is to study the properties of the richest clusters 
of galaxies in the SGW.  Our clusters come from all the superclusters in the SGW so 
that we can compare the dynamical state and galaxy content of the richest 
clusters in different environments, from the high-density core of the richest 
supercluster to clusters in a poor supercluster in the extension of the SGW.  

The present-day dynamical state of clusters of galaxies depends on their 
formation history. One indicator of the former or ongoing mergers between groups 
and clusters is the presence of substructure in clusters
\citep[][and references therein]{2000A&A...354..761K}. The substructure affects 
the estimates of several cluster characteristics, the dynamical mass and 
mass-to-light ratio among others \citep[see][for a review]{2006A&A...456...23B}. 
Mergers also affect the properties of galaxies 
in clusters. The well-known morphology density relation tells that early type, 
red galaxies are located in clusters (in the central areas) while late type, 
blue galaxies can preferentially be found outside of rich clusters, or in the 
outskirts of clusters \citep{1974Natur.252..111E,1980ApJ...236..351D, 
1978ApJ...219...18B,1987MNRAS.226..543E}. An old question is whether the 
properties of galaxies depend on the clustercentric radius or on the local 
density of galaxies in clusters, or on both \citep{1991ApJ...367...64W,
2009A&A...505...83H,2009ApJ...691.1828P,2009ApJ...699.1595P}. The study of
substructure in clusters and of galaxy populations in different substructures 
helps us to understand the role  of 
environmental effects on galaxy populations in clusters.

\begin{figure*}
\centering
\resizebox{0.95\textwidth}{!}{\includegraphics*{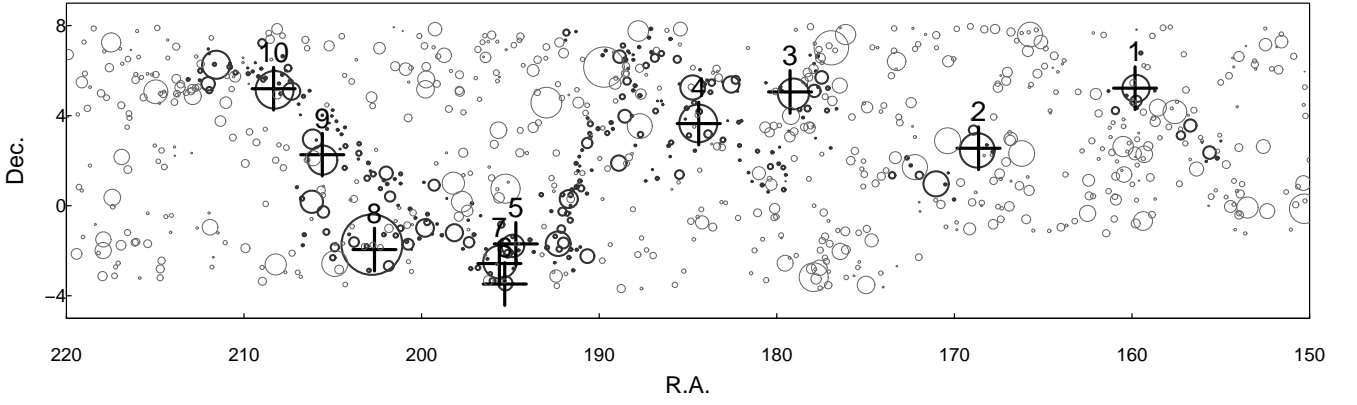}}\\
\caption{
Distribution of groups with at least four member galaxies 
in the Sloan Great Wall region. Black circles correspond to the groups
in the SGW, grey circles to the groups in the field.
Circle sizes are proportional to a group's 
size in the sky. Crosses  show the location of rich clusters, 
and numbers are their ID numbers from Table~\ref{tab:grdata}. 
  }
\label{fig:1}
\end{figure*}

To search for substructure in clusters we use the {\it Mclust} package 
\citep{fraley2006} from {\it R}, an open-source free statistical environment 
developed under the GNU GPL \citep[][\texttt{http://www.r-project.org}]{ig96}. 
One of our aims in this paper is to explore the possibilities of {\it Mclust} 
for the study of substructure in galaxy clusters.

Another indicator of substructure in clusters is the deviation of the 
distribution of the peculiar velocities of galaxies in clusters (the velocities 
of galaxies with respect to the cluster centre) from Gaussian. We use several 
tests to analyse this distribution. 

In virialized clusters the galaxies follow the cluster potential well. If so, we 
would expect that the first ranked galaxies in clusters lie at the centres of 
groups (group haloes) and have low peculiar velocities 
\citep{ost75,merritt84,malumuth92}. Therefore the peculiar velocity  of the 
first ranked galaxies in clusters is also an indication of the 
dynamical state of the cluster \citep{coziol09}.  

Thus, in the present paper we search for subclusters in rich clusters of the SGW, 
analyse the peculiar velocities of galaxies in clusters, and the peculiar 
velocities of the first ranked galaxies. We study the galaxy content of 
substructures, using information about colours, luminosities, and morphological 
types of galaxies.

We compare our results with an N-body simulation and show halo merger trees to 
demonstrate the formation of present-day haloes via multiple mergers during 
their evolution.

\section{Data}

We use the data from the seventh data release
of the Sloan Digital Sky Survey \citep{ade08,aba2009}. We choose 
a subsample of these data from the SGW region: 
$150^\circ\le$R.A.$\le220^\circ$, 
$-4^\circ\le\delta\le 8^\circ$, 
with the distance limits $150\le D_c\le300$\Mpc. 
This region fully 
covers the SGW and also includes several poor superclusters in the foreground and
background of the SGW, 
filaments which connect the SGW with these superclusters, and underdense regions 
between superclusters and filaments. In this region there are 27113 galaxies 
in the MAIN SDSS galaxy sample.

Our next step is to determine groups of galaxies. We next describe 
the sample selection 
and the compilation of the group catalogue; for details we refer to
 \citet{2010A&A...514A.102T}. 
After correction of the galaxy magnitudes 
for galactic extinction 
we use the data about galaxies with the apparent $r$ magnitudes $12.5 \leq r \leq 
17.77$.  In addition,
we have found duplicates due to repeated spectroscopy for a number of galaxies 
in the DAS Main galaxy sample, which had to be excluded in order to avoid false group
members. We count as duplicate entries the 
galaxies which have a projected separation of less than 5 kpc. 
We excluded from our sample those duplicate 
entries with spectra of lower accuracy.  
If both duplicates had a large value of the  redshift confidence parameter 
(zconf $> 0.99$), we excluded the fainter duplicate. 

We corrected the redshifts of galaxies for the motion 
relative to the CMB and computed the  co-moving distances $D_c$ \citep{mar03} of galaxies 
using the standard cosmological parameters: 
the matter density $\Omega_m = 0.27$, and the dark energy density 
$\Omega_{\Lambda} = 0.73$. 

We determined groups of galaxies using the friends-of-friends cluster analysis 
method introduced by \citet{tg76,zes82,hg82}, and modified by us. A galaxy 
belongs to a group of galaxies if this galaxy has at least one group member 
galaxy closer than a linking length. 
In a flux-limited sample the density of galaxies slowly decreases with
distance. 
To take this selection effect properly into account 
when constructing a group catalogue from a flux-limited sample, we 
rescaled the linking length  with distance. As a result, the maximum sizes in 
the sky projection and velocity dispersions of our groups are similar at all 
distances. This shows that the distance-dependent selection effects have been
 properly taken into account, and that groups in our catalogue form a homogeneous
sample suitable for statistical studies. However, there are exceptions:
data about extremely poor groups with less than four member galaxies are not
reliable. In addition, there are four exceptionally rich systems in our
catalogue with more than 300 member galaxies,
which all correspond to well-known nearby Abell clusters
(A1656 (Coma), A2151/2152 (Hercules), A2197/2199 and A1367. 
As we see below, the clusters 
analysed in the present study do not belong 
to these exceptions.

In the group catalogue the first ranked galaxy of a group is defined
as the most luminous galaxy in the $r$-band. We use this definition also
in the present paper. 

The T10 group catalogue is available in electronic form at
the CDS via anonymous ftp to cdsarc.u-strasbg.fr (130.79.128.5) or
via \texttt{http://cdsweb.u-strasbg.fr/cgi-bin/qcat?J/A+A/}.

To select the galaxies and galaxy systems belonging to the SGW the next step was 
to calculate the luminosity density field of galaxies.

To calculate a density field, at first we convert the spatial positions of 
galaxies into spatial densities. The standard approach for that is to use kernel 
densities; we use the $B_3$ box spline

\[
    B_3(x) = \frac{1}{12} \left(|x-2|^3 - 4|x-1|^3 + 6|x|^3 + 4|x+1|^3 + |x+2|^3\right).
\]
We define the $B_3$ box spline kernel of a width $h$ as

\[
    K_B^{(1)}(x;h)_\delta = B_3(x/h) (\delta / h),
\]
where $\delta$ is the grid step. This kernel differs from zero only
in the interval $x\in[-2h;2h]$.
The three-dimensional kernel function $K_B^{(3)}$
is given by the direct product of three one-dimensional kernels:

\[
    K_B^{(3)}(\mathbf{r};h)_\delta \equiv K_3^{(1)}(x;h)_\delta K_3^{(1)}(y;h)_\delta K_3^{(1)}(z;h)_\delta,
\]
where $\mathbf{r} \equiv \{x,y,z\}$. Although this is a direct product,
it is isotropic to a good degree. For a detailed description we refer 
to  \citet{e07}, E10 and 
Liivam\"agi et al. 2010 (in preparation). 

The $k$-correction for the SDSS galaxies was calculated with the KCORRECT
algorithm \citep{blanton03a,bla07}. We also
accepted $M_{\odot} = 4.53$ (in the $r$ photometric system).
The evolution correction $e$ was found according to \citet{blanton03b}.
With the velocity and radial dispersions from the group 
catalogue  by T10 we supressed the cluster-finger redshift distortions.

We also have to consider
the luminosities of the galaxies that lie outside the observational
window of the survey. Assuming that every galaxy is a visible member of a
density enhancement (a group or cluster), we estimate
the amount of unobserved luminosity and weigh each galaxy accordingly:
\begin{equation}
    L_{w} = W_L(d)\; L_{\mathrm{obs}}.
\end{equation}
Here $W_L(d)$ is a distance-dependent weight:
\begin{equation}
    W_L(d) = \frac{\int_0^\infty L\;F(L)dL}{\int_{L_1(d)}^{L_2(d)} L\;F(L) dL},
\end{equation}
where $F(L)$ is the luminosity function and $L_1(d)$ and $L_2(d)$ are the
luminosity window limits at a distance $d$.

In our final flux-limited group  catalogue the richness of groups rapidly decreases 
at distances $D > 300$\Mpc\ due to selection effects. 
  At small distances, $D < 100$\Mpc, the luminosity weights are large 
 owing to the absence of very bright galaxies.
At the distance limits  used to define our sample the selection
effects are small.

The densities were calculated on a cartesian grid based on the SDSS $\eta$, 
$\lambda$ coordinate system because it allowed the most efficient placing of the 
galaxy sample cone into a box. We choose the kernel width $h=8$\Mpc. This kernel 
differs from zero within the distance 16\Mpc, but significantly so only inside 
the 8\Mpc\ radius. We calculated the $x$, $y$, and $z$ coordinates using the $\lambda$ and 
$\eta$ coordinates as follows: $x = -R \sin\lambda$, $y = R \cos\lambda 
\cos\eta$, and $z = R \cos\lambda \sin\eta$, where $R$ is the distance of the 
galaxy. Before extracting superclusters we applied the DR7 mask assembled by 
Arnalte-Mur (\citep{martinez09}  to the density field and converted the densities into 
units of mean density. The mean density is the average over all pixel values 
inside the mask. The mask is designed to follow the edges of the survey, and the 
galaxy distribution inside the mask is considered homogeneous. The details of 
this procedure will be given in Liivam\"agi et al., 2010 (in preparation).

Next we created a set of density contours by choosing a density threshold and 
defined the connected volumes above a certain density threshold as superclusters. 
Different threshold densities correspond to different supercluster catalogues. 
In order to choose proper density levels to determine the SGW and the individual 
superclusters which belong to the SGW, we analysed the  density field superclusters 
at a series of density levels. As a result we used  the density level $D = 4.9$ 
to determine the individual superclusters in the SGW. 

The richest superclusters in the SGW are the superclusters SCl~126 and the 
supercluster SCl~111 (we use the ID numbers of the superclusters from the catalogue of 
superclusters by \citet{e2001}). The supercluster SCl~126 also includes the 
supercluster SCl~136 from the \citet{e2001} catalogue. The core of the supercluster 
SCl~126 contains several  rich X-ray clusters of galaxies \citep{belsole04,e07}. 
The supercluster SCl~126 is the richest in the SGW with a very high density 
core, its morphology resembles a very rich and high-density multibranching 
filament \citep[][E10]{e07,e08}. The supercluster SCl~111 is the second in 
richness in the SGW and consists of three concentrations of rich clusters 
connected by filaments of galaxies \citep[a ``multispider'' morphology, 
see][]{e07}. Poor superclusters from the low-density extension of the SGW belong 
to the supercluster SCl~91 in the \citet{e2001} catalogue. These superclusters 
represent different global environments in the SGW. For details about 
individual superclusters in the SGW we refer to E10.

\begin{figure*}[ht]
\centering
{\resizebox{0.22\textwidth}{!}{\includegraphics*{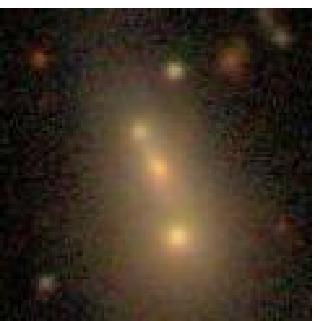}}}
{\resizebox{0.22\textwidth}{!}{\includegraphics*{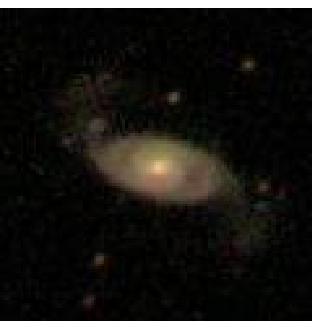}}}
{\resizebox{0.22\textwidth}{!}{\includegraphics*{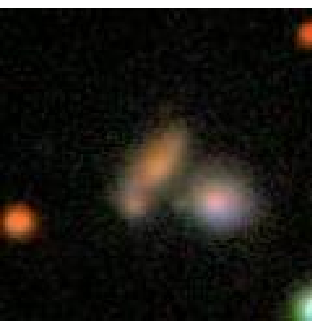}}}
{\resizebox{0.22\textwidth}{!}{\includegraphics*{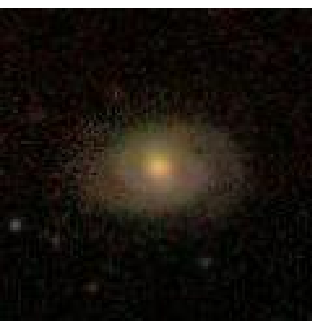}}}
\hspace*{2mm}\\
\caption{
Images of some galaxies in clusters. From left to right: 
the first ranked galaxy in the cluster A1750 with the colour index $g - r = 0.72$,
(J131236.98-011151, BRG), a red spiral galaxy in the cluster A1066
(J10351.33+051130.8, $g - r = 0.75$, BRG), two galaxies in the cluster
A1658 (J130116.68-032823.6 with $g - r = 0.67$, BRG, and J130116.09-032829.1 with 
$g - r = 0.54$), and a red spiral galaxy in the cluster A1663
(J130152.6-024012.3 with $g - r = 0.75$, BRG).
}
\label{fig:images}
\end{figure*}

\begin{table*}[ht]
\caption{
Rich clusters in the SGW.
}
\begin{tabular}{rrrrrrrrrrr} 
\hline 
(1)&(2)&(3)&(4)& (5)&(6)&(7)& (8)&(9)& (10)&(11)\\      
\hline 
Scl ID  &No& ID  & Dist. & $R.A.$ &$Dec.$ & L$_{\mbox{tot}}$& $\sigma_{\mbox{v}}$& N$_{\mbox{gal}}$ & N$_{\mbox{BRG}}$ &N$_{\mbox{BRS}}$\\        
        &  &     &[$h^{-1}$ Mpc]&[deg]&[deg]&$10^{10} h^{-2} L_{\sun}$& $km s^{-1}$&   &&\            
\\
SCl~91  &1& 1066    & 206.6 & 159.8 & 5.2 & 0.854E+02 & 758 &   78 & 15 &  8 \\
        &2& 1205    & 228.4 & 168.6 & 2.3 & 0.108E+03 & 624 &  105 & 11 &  1 \\
SCl~111 &3& 1424    & 226.0 & 179.3 & 5.1 & 0.968E+02 & 656 &   78 & 12 &  4  \\
        &4& 1516    & 230.3 & 184.4 & 3.7 & 0.119E+03 & 834 &  104 & 15 &  3  \\ 
SCl~126 &5& 1650x   & 249.8  & 194.7 & -1.7 & 0.706E+02 & 466 & 48 &  8 & 3 \\
        &6& 1658x   & 253.4  & 195.3 & -3.5 & 0.445E+02 & 441 & 29 &  4 & 1 \\
        &7& 1663x   & 246.7  & 195.6 & -2.6 & 0.104E+03 & 530 & 78 & 12 & 2 \\
        &8& 1750x   & 256.4  & 202.7 & -1.9 & 0.190E+03 & 660 &117 & 17 & 2 \\
        &9& 1773x   & 230.6  & 205.6 &  2.3 & 0.918E+02 & 646 & 74 & 11 & 4  \\
        &10&1809    & 236.5  & 208.3 &  5.2 & 0.118E+03 & 649 & 90 & 12 & 3  \\
\

\label{tab:grdata}                        
\end{tabular}\\
\small\rm\noindent Columns in the Table are as follows:
\noindent Cols. 1--3: Supercluster ID, cluster No and Abell ID (x denotes X-ray clusters),
Cols. 4--6: comoving distance, R.A. and Dec. of the cluster, 
Col. 7: total luminosity of the supercluster L$_{\mbox{tot}}$, 
Col. 8: $\sigma_{\mbox{v}}$ - rms velocity,
Cols. 9--11: N$_{\mbox{gal}}$ -- number of galaxies,
N$_{\mbox{BRG}}$ -- number of bright red galaxies,
N$_{\mbox{BRS}}$ -- number of spirals among the bright red galaxies.
\end{table*}

Then we selected  rich clusters from the SGW for our study 
as follows. First, we selected  all
clusters of galaxies in the SGW with at least 75 member galaxies,  altogether 7 
clusters. We added to this sample a cluster which corresponds to the Abell 
cluster A1773, an X-ray cluster \citep{boe04}. This cluster has 74 member 
galaxies in our catalogue. We also added to our analysis two clusters from the 
core region of the supercluster SCl~126, A1650 with 48 member galaxies and A1658 
with 29 member galaxies. Both of them are X-ray clusters \citep{boe04}. The 
reason is that we wanted to study in detail all rich clusters in the core region 
of this supercluster. Our earlier analysis 
(E10) has shown that there are several differences in the properties 
of groups and of the galaxy content between  the core of the supercluster
SCl~126 and other superclusters in the SGW: groups in the core of
the supercluster SCl~126 are richer than in other superclusters 
of the SGW, and
the fraction of red galaxies there is larger than in the outskirts of this 
supercluster or in other superclusters in the SGW.  Now we compared the 
properties of rich clusters in the core of the supercluster SCl~126 with the 
properties of rich clusters from other superclusters in the SGW  to see 
whether the dynamical state of clusters differs.

Thus, our sample of rich clusters consists of ten clusters and includes all
very rich clusters in the SGW, all X-ray clusters and all rich clusters from the 
core of the supercluster SCl~126. All these clusters correspond to Abell 
clusters; their Abell ID and other data are given in Table~\ref{tab:grdata}. We 
show the sky distribution of groups in the SGW in Fig.~\ref{fig:1}.
In this figure clusters are marked with numbers from 
Table~\ref{tab:grdata}.  The clusters 1 and 2 are located in the supercluster 
SCl~91, the clusters  3 and 4  in the supercluster SCl~111, the clusters 5--8 
 in the core of the supercluster SCl~126 and the clusters 9 and 10 in the 
outskirts of the supercluster SCl~126. 

Our clusters are chosen from a narrow distance interval
(Table~\ref{tab:grdata}), here our sample is almost volume-limited.
Absolute magnitude limits of galaxies in groups in the superclusters SCl~91
and  SCl~111 are approximately -19.20, and in the supercluster SCl~126 
-19.40.

\section{Methods}
\subsection{Substructure in rich clusters}

\begin{table*}[ht]
\caption{Dynamical properties of rich clusters.
}
\begin{tabular}{rrrrrrrrrrrr} 
\hline 
(1)&(2)&(3)&(4)& (5)&(6)& (7)&(8) &(9) &(10)&(11)  \\      
\hline 
Scl ID  &No& ID  &    N$_{\mbox{comp}}$& $v_{\mbox{pec}}$ &$v_{\mbox{rpec}}$ &$p_{\mbox{SW}}$ &k&$p_{\mbox{k}}$&sk&$p_{\mbox{sk}}$\\
\\          
SCl~91  &1& 1066      & 1 &-312 & 0.41 & 0.074  & 2.34 & 0.139 & -0.243 &  0.528  \\
        &2& 1205      & 5 &  49 & 0.08 & 0.247  & 2.37 & 0.099 & -0.069 &  0.841  \\
SCl~111 &3& 1424      & 2 &-825 & 1.26 & 0.173  & 2.51 & 0.387 &  0.383 &  0.339 \\
        &4& 1516      & 1 & -58 & 0.07 & 0.535  & 2.84 & 0.948 & -0.292 &  0.403 \\ 
SCl~126 &5& 1650x     & 1 &-189 & 0.41 & 0.110  & 2.80 & 0.946 &  0.563 &  0.263 \\
        &6& 1658x     & 1 &-463 & 1.05 & 0.937  & 2.37 & 0.583 &  0.048 &  0.936 \\
        &7& 1663x     & 2 &-841 & 1.59 & 0.372  & 2.57 & 0.493 & -0.228 &  0.562 \\
        &8& 1750x     & 5 &-724 & 1.10 & 0.115  & 2.51 & 0.243 & -0.265 &  0.422 \\
        &9& 1773x     & 1 & -76 & 0.12 & 0.572  & 3.24 & 0.436 & -0.083 &  0.835 \\
        &10&1809      & 2 &-115 & 0.18 & 0.802  & 2.73 & 0.780 &  0.053 &  0.887 \\
\\

\label{tab:grdyn}                        
\end{tabular}\\
\small\rm\noindent Columns in the Table are as follows:
\noindent 1--3: Supercluster ID, cluster No and Abell ID,
Col. 4: number of components,
Col. 5: peculiar velocity of the first ranked galaxy
($km s^{-1}$),
Col. 6: normalised peculiar velocity of the first ranked galaxy,
$v_{\mbox{pec}}$/$\sigma_{\mbox{v}}$;
Col. 7:
$p$-value of the Shapiro-Wilk test,
Cols. 8 and 9: kurtosis $k$ and the $p$-value
$p_{\mbox{k}}$  for kurtosis for the peculiar velocities, 
Cols. 10 and 11: skewness $sk$ and the $p$-value
 $p_{\mbox{sk}}$  for skewness for the peculiar velocities.
\end{table*}

\begin{table*}[ht]
\caption{Galaxy populations in rich clusters.
}
\begin{tabular}{rrrrrrrrrrr} 
\hline 
(1)&(2)&(3)&(4)& (5)&(6)&(7)& (8)& (9)&(10)&(11)   \\      
\hline 
Scl ID  &No& ID  &   $(g-r)_{\mbox{mean}}$& F$_{\mbox{red}}$ &  F$_{\mbox{ell}}$ & F$_{\mbox{rs}}$ &slope$_{\mbox{red}}$ & SD$_{\mbox{red}}$&SD$_{\mbox{Ered}}$&SD$_{\mbox{Sred}}$ \\        
\\                         
SCl~91  &1& 1066     &  0.73 & 0.73 & 0.36 & 0.58 & -0.024 & 0.043  & 0.024 & 0.053 \\
        &2& 1205     &  0.68 & 0.62 & 0.38 & 0.45 & -0.012 & 0.030  & 0.027 & 0.033 \\
SCl~111 &3& 1424     &  0.73 & 0.71 & 0.44 & 0.49 & -0.007 & 0.046  & 0.033 & 0.057 \\
        &4& 1516     &  0.76 & 0.85 & 0.56 & 0.34 & -0.018 & 0.032  & 0.032 & 0.033 \\ 
SCl~126 &5& 1650x     & 0.78 & 0.94 & 0.56 & 0.40 & -0.010 & 0.039  & 0.036 & 0.042 \\
        &6& 1658x     & 0.74 & 0.83 & 0.58 & 0.29 & -0.026 & 0.029  & 0.031 & 0.024 \\
        &7& 1663x     & 0.75 & 0.83 & 0.49 & 0.43 & -0.007 & 0.037  & 0.032 & 0.042 \\
        &8& 1750x     & 0.78 & 0.91 & 0.62 & 0.33 & -0.013 & 0.042  & 0.041 & 0.046 \\
        &9& 1773x     & 0.73 & 0.77 & 0.53 & 0.41 & -0.011 & 0.040  & 0.032 & 0.048 \\
        &10&1809      & 0.75 & 0.84 & 0.51 & 0.43 & -0.015 & 0.029  & 0.027 & 0.032 \\
\\

\label{tab:grgal}                        
\end{tabular}\\
\small\rm\noindent Columns in the Table are as follows:
\noindent Cols. 1--3: Supercluster ID, cluster No and Abell ID,
Col. 4:   $(g - r)_{\mbox{mean}}$ -- the mean $g - r$ colour of galaxies, 
Cols. 5--11: fraction of red galaxies, fraction of elliptical 
galaxies, fraction of spirals among red galaxies, and the slope and the rms scatter
of the red sequence in the colour -- magnitude diagram (the scatter of all
red galaxies, and that of red ellipticals and red spirals).
\end{table*}

We studied the substructure of clusters with  the $R$ package
{\it Mclust}\citep{fraley2006}.
This package searches for an optimal model for the  clustering of 
the data among the  models with varying shape, orientation, and volume, and for 
the optimal number of components or substructures, using multidimensional normal mixture
modeling. 
Below we will use {\it Mclust} to calculate the number of 
components in clusters using the data about the sky position and peculiar velocities 
of the cluster member galaxies. To see how these components are populated 
with galaxies of different colour and luminosity we applied {\it Mclust} to an extended 
dataset including the colours and luminosities of the cluster members. 
{\it Mclust} calculates for every galaxy the probabilities to belong to any of the components.
The uncertainty of classification is defined as 
1. minus the highest probability of a galaxy to belong to a component.
The mean uncertainty for the sample is used  as a statistical estimate of the reliability of the results.

When we searched for substructure in 3D, we assumed that the line-of-sight 
positions of galaxies are given by their peculiar velocities, because we had no other 
hypothesis that could be used. However, while peculiar velocities might carry 
some distance information, the only fact we can safely assume is that the 
cluster galaxies are located inside the cluster volume. This is the reason why we 
studied the velocity distribution separately. We also tested how the possible 
errors in the line-of-sight positions of galaxies affect the results  of {\it 
Mclust}  randomly  shuffling the peculiar velocities of galaxies 1000 times and 
searching  each time for the substructure with {\it Mclust}. The number of the 
components found by {\it Mclust} remained unchanged, which demonstrated that the 
results of {\it Mclust} are robust.

To analyse the distribution of the peculiar velocities of galaxies in clusters 
we used several 1D tests. We tested the hypothesis about the Gaussian 
distribution of the peculiar velocities of galaxies in clusters with the 
Shapiro-Wilk normality test \citep{shapiro65}. We used this test because it is 
considered the best for small samples. We also calculated the kurtosis and the 
skewness of the peculiar velocity distributions, and used these to test for the 
normality of the distributions with the Anscombe-Glynn test for the kurtosis 
\citep{anscombe83} and with the D'Agostino test for the skewness 
\citep{agostino70}, using the $R$ package {\it moments} by L. Komsta and F. 
Novomestky (\texttt{http://www.r-project.org}).

The distribution of the peculiar velocities of galaxies in clusters can be 
quantified with the usual normal mixture models (e.g. 
\citet{1994AJ....108.2348A,2006A&A...449..461B, 2007A&A...469..861B}. In the $R$ 
environment there is a suitable  package to study these mixture models, called {\it 
flexmix} \citep{leisch2004}. {\it Flexmix} fits a user-specified number of 
Gaussians to the velocity data. However, subclusters in clusters sometimes do 
not differ in peculiar velocities, and in some clusters the tails of the 
velocity distribution are not well determined by {\it flexmix}. Thus we fitted 
the distributions of the peculiar velocities with normal mixture models 
ourselves, taking into account the substructure information provided by {\it 
Mclust} and by the 3D distribution of galaxies that we will show below.

The data characterizing the clusters' dynamical state are given in 
Table~\ref{tab:grdyn}, where we present the number of components in clusters 
determined by {\it Mclust}, the peculiar velocities of the first ranked galaxies, 
the results of the Shapiro-Wilk test, 
the kurtosis and skewness, and their normality test $p$-values for the 
peculiar velocity distributions.

\subsection{Galaxy populations in rich clusters}

To analyse the galaxy content of clusters, we used  the $g - r$ colour of galaxies and 
their absolute magnitude in the $r$-band $M_r$. We divided the galaxies by colour 
into the red and blue populations using the colour limit $ g - r = 0.7$ (red galaxies 
have $ g - r  \geq 0.7$) and excluded from the analysis the galaxies with $g -
r > 0.95$ (such a high value indicates possible errors in photometry, often due to 
overlapping images of galaxies). This limit depends on the luminosity of 
galaxies; because there are no very faint galaxies in our sample, we will use this 
simple approach. We calculated the fraction of red galaxies in clusters, the  mean 
colour of the cluster, and the rms scatter and  slope of the red sequence in the 
cluster colour-magnitude diagram (for all galaxies, for red 
ellipticals, and for red spirals).

We also studied the population of bright red galaxies (BRGs, the galaxies that 
have the GALAXY\_RED flag in the SDSS database) in the clusters. The BRGs are 
nearby ($z < 0.15$, cut I) LRGs \citep{eisenstein01}. Because
\citet{eisenstein01} warn that the sample of nearby LRGs may be contaminated by 
galaxies of lower luminosity, we choose
to call them BRGs.   The BRGs are similar to the LRGs at higher redshifts. 
Nearby bright red galaxies do not form an approximately volume-limited 
population \citep{eisenstein01} but they are yet the most bright and the most 
red galaxies in the SGW region (see also E10).

We determined the morphological type of galaxies in clusters with the  
SDSS visual tools. Following the classification used in the ZOO project 
\citep{lintott08}, we denote elliptical galaxies as of type 1, 
spiral galaxies of type 4, 
those galaxies which show a signs of merging or other disturbancies as of type 
6. Types 2 and 3 mark clockwise and anti-clockwise spiral galaxies with visible
spiral arms. Fainter galaxies, for which the morphological type is 
difficult to determine, are of type 5. We calculate the fraction of elliptical
galaxies and of red spirals in clusters.  A word of caution is needed: 
faint galaxies are sometimes difficult to classify 
reliably. Some red galaxies classified as spirals may actually
be S0 galaxies. We present images of some galaxies in
clusters in Fig.~\ref{fig:images}. All these galaxies are classified
as BRGs (an exception is a blue galaxy in the third image from the left,
which may be interacting with a companion galaxy).

The data for cluster galaxy populations are given in Table~\ref{tab:grgal}. 
Images of all galaxies in our clusters are shown on 
the web page \texttt{http://www.aai.ee/$\sim$maret/SGWcl.html},

In the following analysis of individual rich clusters we present for each 
cluster the sky distribution of galaxies along with the density contours, the 3D 
vizualisation plot of galaxies  (to generate these figures, we used the $R$ 
package {\it scatterplot3D} by \citet{liggesmachler2003}), the histograms of the 
peculiar velocities for galaxies of a different morphological type and for the 
BRGs, and the colour-magnitude diagrams. We have added the Gaussians for the components
to the histograms of the 
peculiar velocities of all galaxies. For some 
clusters we include additional figures to show their substructure.

In the figures of the sky distribution of galaxies in clusters the 2D density 
contours are calculated with {\it Mclust} for clusters  with more  than one component, 
and with the $R $package {\it Kernsmooth}  for clusters  with one component only. 

\section{Results}

\subsection{Clusters in the superclusters SCl~111 and SCl~91}

\subsubsection{The cluster A1066}

\begin{figure}
\centering
{\resizebox{0.22\textwidth}{!}{\includegraphics*{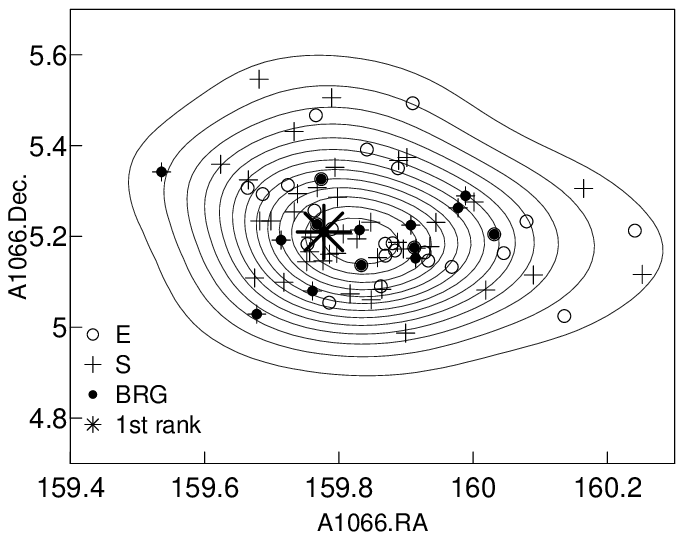}}}
{\resizebox{0.25\textwidth}{!}{\includegraphics*{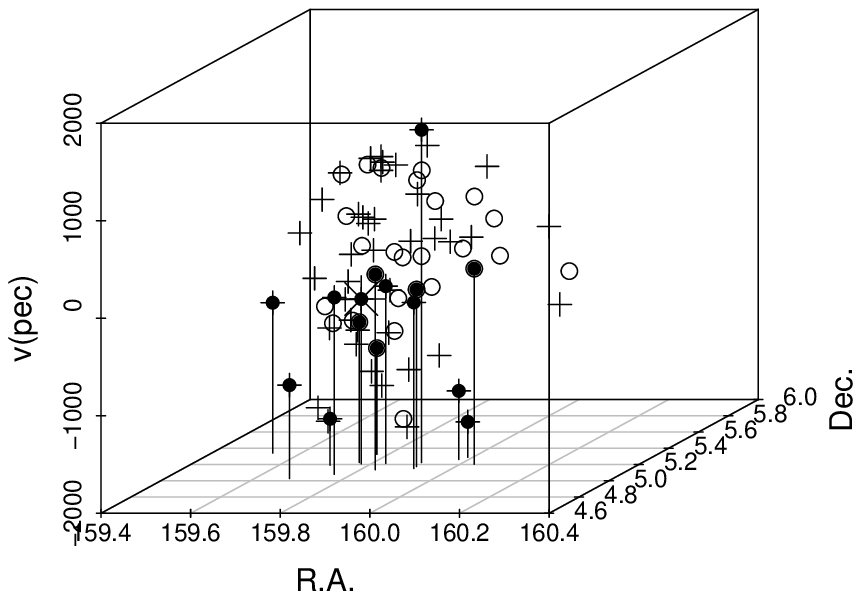}}}
\hspace*{2mm}\\
\caption{2D and 3D  view of the cluster A1066 (SCl~91).
Open circles correspond to  elliptical galaxies, crosses -- to spiral galaxies,
filled circles to BRGs, and a star to the first ranked galaxy in the cluster.
}
\label{fig:gr225441}
\end{figure}

The first rich cluster we study is the cluster A1066, which is located 
in the supercluster SCl~91. Figure~\ref{fig:gr225441} (left panel) shows that the 2D 
density contours of this cluster are quite smooth ellipsoids. According to 
{\it Mclust},  this cluster consists of one component, the mean 
uncertainty of the classification is $1.89\cdot10^{-2}$. However, the 3D distribution of 
galaxies in this cluster Fig.~\ref{fig:gr225441} (right panel) is complex -- 
the shape resembles an hourglass, where some galaxies with 
negative peculiar velocities form a separate component.

\begin{figure}
\centering
{\resizebox{0.45\textwidth}{!}{\includegraphics*{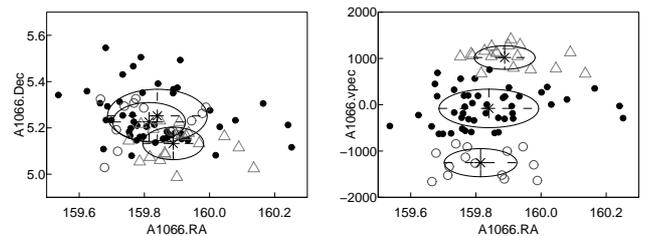}}}
\hspace*{2mm}\\
\caption{ 
Left panel shows  R.A. vs. Dec., the right panel  R.A. vs. 
peculiar velocities of galaxies (in $km s^{-1}$)  in the cluster A1066. 
Different symbols correspond to three different velocity components
(see text), ellipses show the covariances of the components.
}
\label{fig:gr225444}
\end{figure}

\begin{figure}
\centering
{\resizebox{0.30\textwidth}{!}{\includegraphics*{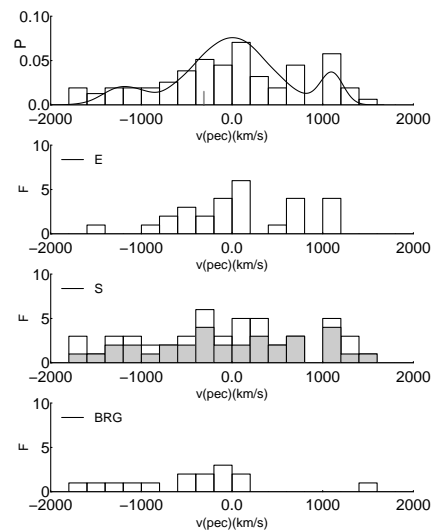}}}
\hspace*{2mm}\\
\caption{Distribution of the
peculiar velocities of galaxies in the cluster A1066. Top down: all galaxies,
elliptical galaxies, spiral galaxies (the grey histogram shows
the red spirals), and BRGs.
P denotes the probability histogram, F the galaxy number counts.
The small dash in the upper panel indicates the peculiar velocity of the first 
ranked galaxy.
}
\label{fig:gr225442}
\end{figure}

Figure~\ref{fig:gr225444}, a so-called classification plot produced by {\it Mclust}, 
shows three components in the distribution of the peculiar velocities of 
galaxies. The peculiar velocities with negative 
values correspond to the lower part of the ``hourglass''. The left panel of this figure 
shows  that the different components in velocity are not separated in the sky 
distribution, which is why {\it Mclust} determines one component only for this 
cluster as given in Table~\ref{tab:grdyn}. We see three 
components in the distribution of the peculiar velocities of galaxies in this 
cluster, and the $p$-value of the Shapiro-Wilk test for this cluster is 0.074, the 
lowest among our sample (the lower the $p$-value, the higher is the difference to
the Gaussian null hypothesis). 

The histograms of the peculiar velocities for all galaxies and separately for the
elliptical and spiral (and red spiral) galaxies and for the BRGs in A1066 are shown in 
Fig.~\ref{fig:gr225442}. At the peculiar velocities $v_{pec} > 1000$km/s we 
see a small separate component of galaxies consisting mostly of spirals 
(including red spirals), although there are also some elliptical galaxies and one 
BRG among them; this system is also seen in Fig.~\ref{fig:gr225444}. In the main 
component of this cluster the galaxies have peculiar velocities $1000 > v_{pec} > -
500$ km/s, the first ranked galaxy of the cluster also belong here. At negative 
peculiar velocities ($v_{pec} \leq - 500$km/s) there is another 
extension (a lower part of the ``hourglass'' in Fig.~\ref{fig:gr225441}, right 
panel) of galaxies, consisting again mostly of spirals, including red 
spirals and five BRGs. Eight of sixteen BRGs in this cluster have been classified 
as spirals (Table~\ref{tab:grgal}), five of them are located in this extension. 
The distribution of the peculiar velocities of elliptical galaxies shows a 
concentration towards the cluster centre, while the distribution of the peculiar 
velocities of spiral galaxies is smoother.

\begin{figure}
\centering
{\resizebox{0.35\textwidth}{!}{\includegraphics*{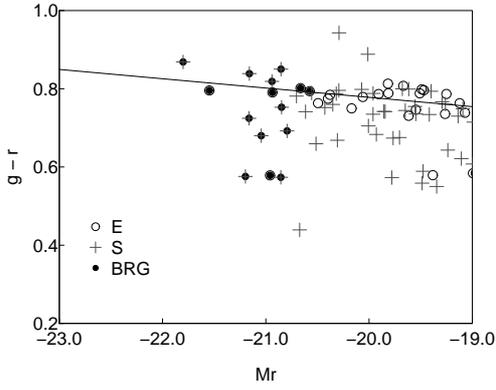}}}
\hspace*{2mm}\\
\caption{Colour-magnitude diagram  of galaxies in the cluster A1066.
Empty circles denote elliptical galaxies, crosses  spiral galaxies,
filled circles BRGs.
}
\label{fig:gr225443}
\end{figure}

In the colour-magnitude diagram for  A1066 (Fig.~\ref{fig:gr225443}) we show 
with separate symbols elliptical and spiral galaxies and BRGs. The slope and the rms
scatter of the red sequence in the colour-magnitude diagram are presented in 
Table~\ref{tab:grgal}.  73\% of the galaxies in this cluster are red, more than 
half of the red galaxies are spirals (or S0s). The colour-magnitude diagram shows 
that the scatter of colours of red spirals (both the red spiral BRGs and fainter red 
spirals) is larger than the scatter of colours of red elliptical galaxies (two 
faint blue galaxies in this cluster are classified as blue ellipticals), which
suggests that red elliptical galaxies in this cluster form a more homogeneous 
population than red spiral galaxies. High negative and high positive 
peculiar velocities of some spiral galaxies and some BRGs in this cluster 
suggest that they dynamically form separate components of the cluster.

Summarizing, this cluster, classified by {\it Mclust} as a one-component 
cluster with a rather smooth sky distribution of galaxies has indeed a 
substructure, which suggests that this cluster has not yet completed its formation.

\subsubsection{The cluster A1205}

\begin{figure}
\centering
{\resizebox{0.22\textwidth}{!}{\includegraphics*{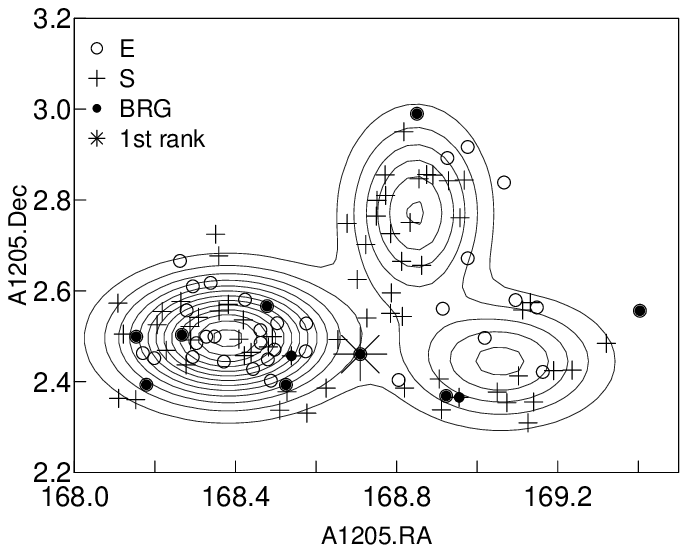}}}
{\resizebox{0.25\textwidth}{!}{\includegraphics*{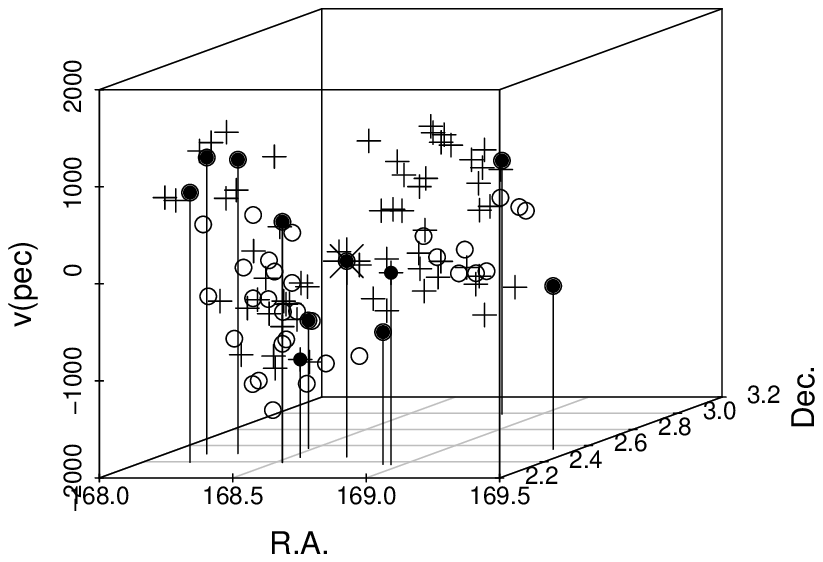}}}
\hspace*{2mm}\\
\caption{2D and 3D views of the cluster A1205 (SCl~91).
Symbols are the same as in Fig.~\ref{fig:gr225441}.
}
\label{fig:gr276811}
\end{figure}

\begin{figure}
\centering
{\resizebox{0.45\textwidth}{!}{\includegraphics*{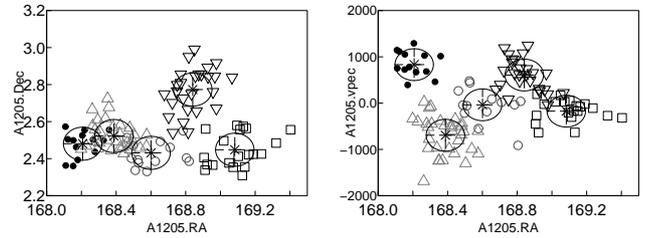}}}
\hspace*{2mm}\\
\caption{
Left panel shows R.A. vs. Dec., the right panel the R.A. vs. the
peculiar velocities of galaxies (in $km s^{-1}$)  in the cluster A1205.
Different symbols correspond to different velocity components. 
}
\label{fig:gr276815}
\end{figure}

The sky distribution of galaxies in the  cluster A1205  in the supercluster 
SCl~91 (Fig.~\ref{fig:gr276811}, left panel) shows  three 
different components. According to  {\it Mclust},  this cluster 
 even consists of five components, the mean uncertainty of the classification is 
$1.62\cdot10^{- 3}$. The reason for that is the distribution of the peculiar velocities 
of galaxies in each component that is shown in Fig.~\ref{fig:gr276815}. This 
figure shows that the galaxies in a left component are divided into three according 
to their peculiar velocities; part of them have high positive peculiar 
velocities and others  mostly negative peculiar velocities. There is also a 
small subcluster of galaxies with very low peculiar velocities.

\begin{figure}
\centering
{\resizebox{0.30\textwidth}{!}{\includegraphics*{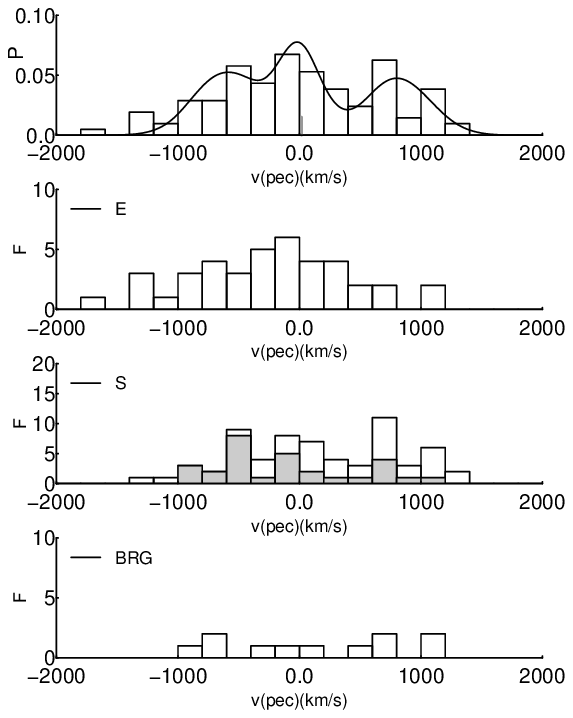}}}
\hspace*{2mm}\\
\caption{Distribution of the
peculiar velocities of galaxies in the cluster A1205. Top down: all galaxies,
elliptical galaxies, spiral galaxies, and BRGs.
The small dash in the upper panel indicates the peculiar velocity of the first 
ranked galaxy.
}
\label{fig:gr276812}
\end{figure}

In another big component the galaxies have mostly positive
peculiar velocities, in the third one -- negative peculiar velocities
(Fig.~\ref{fig:gr276812}). 
The shape of the 3D distribution of galaxies 
Fig.~\ref{fig:gr276811} (right panel)  resembles
an hourglass, this time seen at an angle. 

\begin{figure}
\centering
{\resizebox{0.35\textwidth}{!}{\includegraphics*{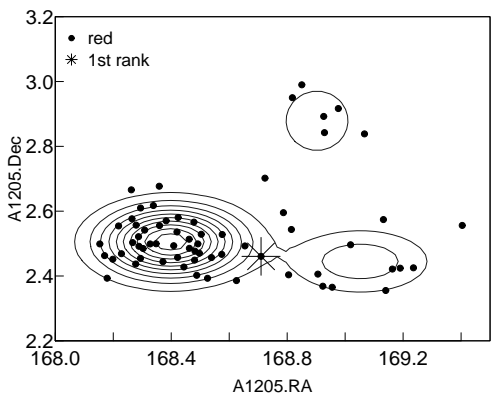}}}
\hspace*{2mm}\\
\caption{Distribution of red galaxies (filled circles) in the cluster A1205. 
The star denotes the first ranked galaxy.
}
\label{fig:gr276814}
\end{figure}

 Even more interesting is the structure of this
cluster if we plot the sky distribution with the density contours for red
galaxies only (Fig.~\ref{fig:gr276814}). This figure shows that one component
in this cluster mostly consists of red galaxies (there are both elliptical
and spiral galaxies among them, compare Fig.~\ref{fig:gr276811}, left panel
and Fig.~\ref{fig:gr276814}). 

The first ranked galaxy in this cluster is located almost at 
the centre of the sky distribution and has a very low peculiar velocity
(the lowest among the clusters in our sample, see Table~\ref{tab:grdyn}).

The histograms of the peculiar velocities of galaxies in A1205 in Fig.~\ref{fig:gr276812} 
show a large fraction of red spirals with negative peculiar velocites, which
belong to the component with red galaxies. Some BRGs are also located there; 
in addition, we see several BRGs in other components, even at the edges of the 
cluster (Fig.~\ref{fig:gr276811}).
The peculiar velocities of galaxies
from the three different components seen in the sky distribution 
partly overlap, so these components are not very clearly seen
in the  histograms.

\begin{figure}
\centering
{\resizebox{0.35\textwidth}{!}{\includegraphics*{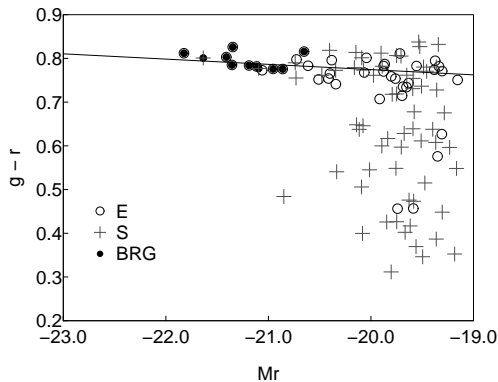}}}
\hspace*{2mm}\\
\caption{Colour-magnitude diagram of galaxies in the cluster A1205.
Empty circles denote elliptical galaxies, crosses spiral galaxies,
filled circles BRGs.
}
\label{fig:gr276813}
\end{figure}

In the colour-magnitude diagram  for the A1205 
(Fig.~\ref{fig:gr276813}) we see a bimodal distribution of galaxies, where the red and 
blue sequences are both clearly seen. This cluster has the largest fraction of blue 
galaxies in our sample (32\%). The colour-magnitude diagram shows that the scatter of 
colours of red spirals  is larger 
than the scatter of colours of red elliptical galaxies (the rms scatters are 0.033 and 0.027, 
correspondingly, for red spirals and for red ellipticals),  although the 
difference is not as pronounced  as in the cluster A1066. In this 
cluster there are four faint blue galaxies classified as blue ellipticals. 

The presence of substructures in  this cluster, which are
populated by galaxies of different properties, suggests that the cluster A1205
consists of at least three merging components.

\subsubsection{The cluster A1424}

\begin{figure}
\centering
{\resizebox{0.22\textwidth}{!}{\includegraphics*{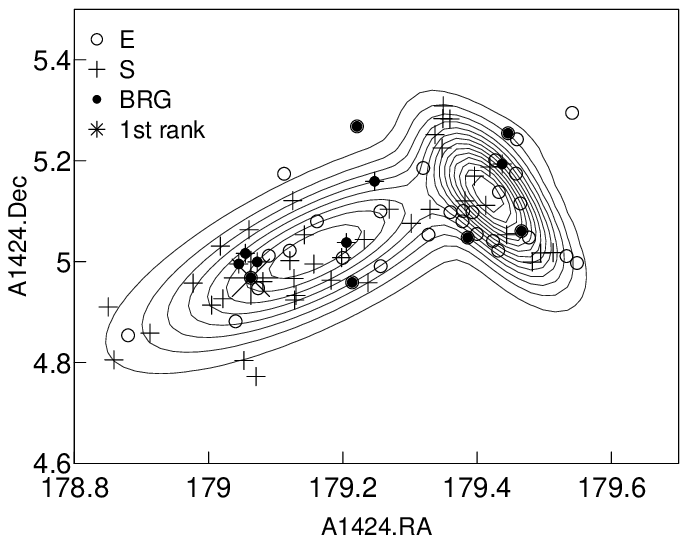}}}
{\resizebox{0.25\textwidth}{!}{\includegraphics*{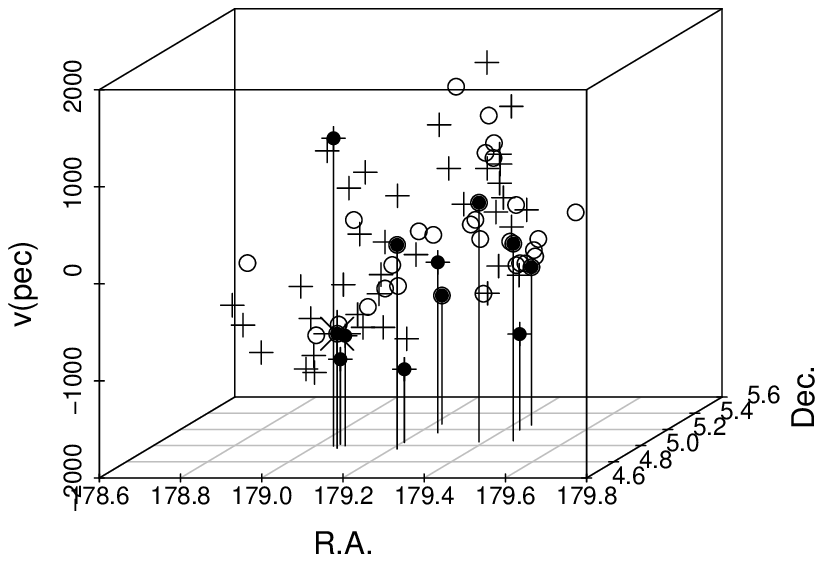}}}
\hspace*{2mm}\\
\caption{2D and 3D views of the cluster A1424 (SCl~111).
Symbols are the same as in Fig.~\ref{fig:gr225441}.
}
\label{fig:gr349991}
\end{figure}

The next cluster under study, the cluster A1424, 
is located in the second rich supercluster in the SGW, SCl~111.
The sky distribution of galaxies in this cluster 
(Fig.~\ref{fig:gr349991}, left panel) shows  two components
also confirmed  by  {\it Mclust}. 
The mean uncertainty of the classification of galaxies in the cluster A1424  
is $2.92\cdot10^{-4}$.

\begin{figure}
\centering
{\resizebox{0.22\textwidth}{!}{\includegraphics*{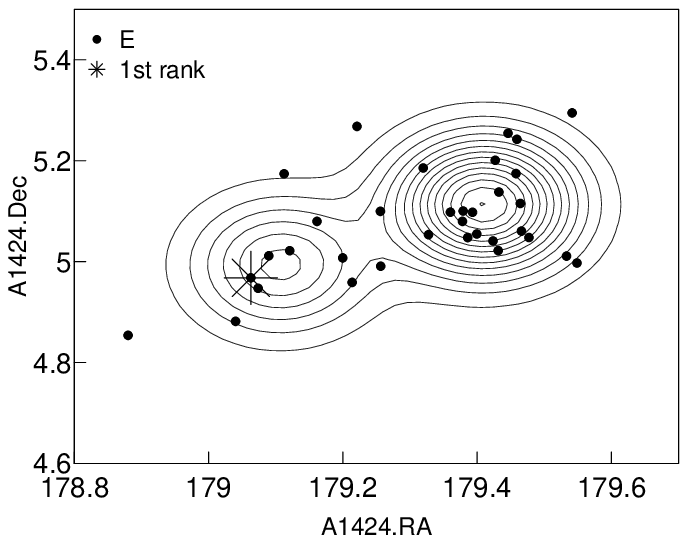}}}
{\resizebox{0.22\textwidth}{!}{\includegraphics*{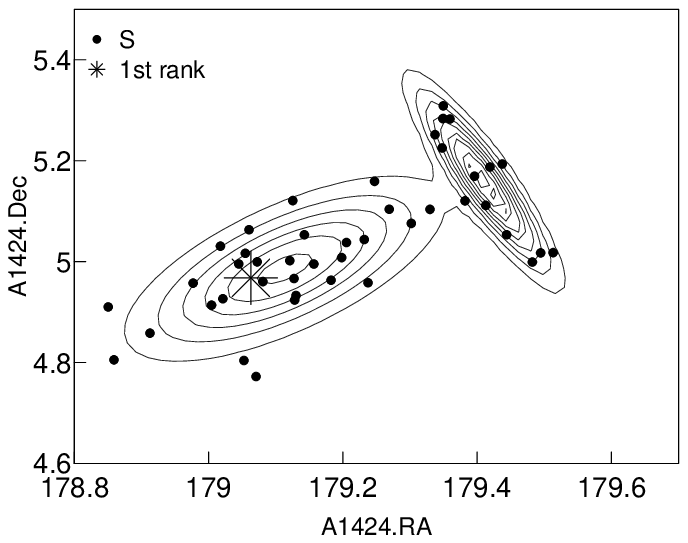}}}
\hspace*{2mm}\\
\caption{Sky distribution of 
elliptical (left) and spiral (right) galaxies in the cluster A1424. 
The 2D density contours are 
calculated using the data about elliptical or spiral
galaxies, respectively. Filled circles in the left panel denote elliptical galaxies,
in the right panel  spiral galaxies. The star denotes the first ranked galaxy.
}
\label{fig:gr349994}
\end{figure}

The 3D distribution of galaxies (Fig.~\ref{fig:gr349991}, right panel)  shows 
that in one of these components galaxies have mostly positive peculiar 
velosities. Figure~\ref{fig:gr349994} shows that according to the sky 
distribution of galaxies in the cluster A1424, the components as delineated by 
elliptical and spiral galaxies are  different -- the two components  
are more clearly separated in the distribution of spiral galaxies.  

\begin{figure}
\centering
{\resizebox{0.30\textwidth}{!}{\includegraphics*{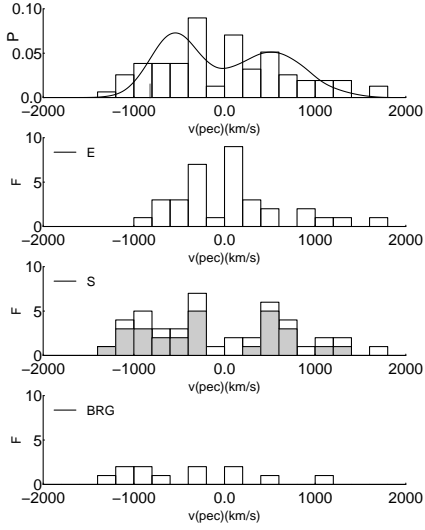}}}
\hspace*{2mm}\\
\caption{Distribution of the
peculiar velocities of galaxies  in the cluster A1424. Top down: all galaxies,
elliptical galaxies, spiral galaxies, and BRGs.
The small dash in the upper panel indicates the peculiar velocity of the first 
ranked galaxy.
}
\label{fig:gr349992}
\end{figure}

The distribution of the peculiar velocities of galaxies in the cluster A1424 
(Fig.~\ref{fig:gr349992}) has a minimum near the
clusters centre, which is seen in the distribution of the peculiar velocities of 
both elliptical and spiral galaxies. The elliptical galaxies are  more 
concentrated towards the centre of the cluster than the spiral galaxies, which is why
the distribution of the peculiar velocities is smoother. In the same time, the 
distribution of the peculiar velocities of red spirals also shows two 
separate components. The BRGs are mostly located in the component with smaller right 
ascensions, several of which are spirals. The first ranked galaxy of this 
cluster is located almost in the centre of this component. 

\begin{figure}
\centering
{\resizebox{0.35\textwidth}{!}{\includegraphics*{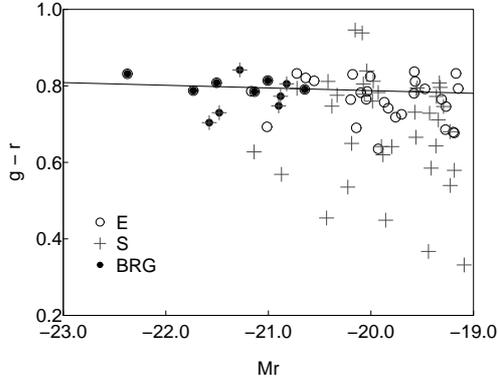}}}
\hspace*{2mm}\\
\caption{Colour-magnitude diagram for galaxies  in the cluster A1424.
Empty circles denote elliptical galaxies, crosses spiral galaxies,
filled circles  BRGs.
}
\label{fig:gr349993}
\end{figure}

In this cluster 73\% of all galaxies 
are red. In the colour-magnitude diagram (Fig.~\ref{fig:gr349993})
the scatter of red spiral galaxies is about two times larger than
the scatter of red elliptical galaxies
(the rms scatters are 0.033 and 0.057, 
correspondingly, for red ellipticals and for red spirals).  
This is caused by galaxies in the outer parts of the cluster; it is possible that
these galaxies have only recently joined the cluster. 

\subsubsection{The cluster A1516}

\begin{figure}
\centering
{\resizebox{0.22\textwidth}{!}{\includegraphics*{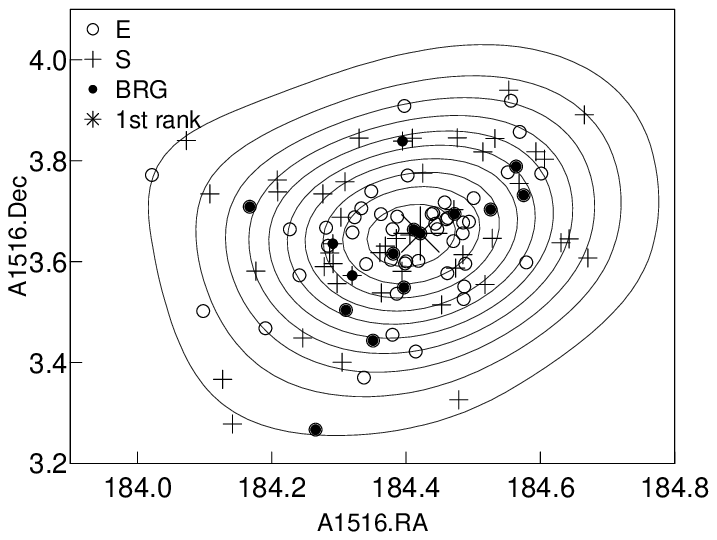}}}
{\resizebox{0.25\textwidth}{!}{\includegraphics*{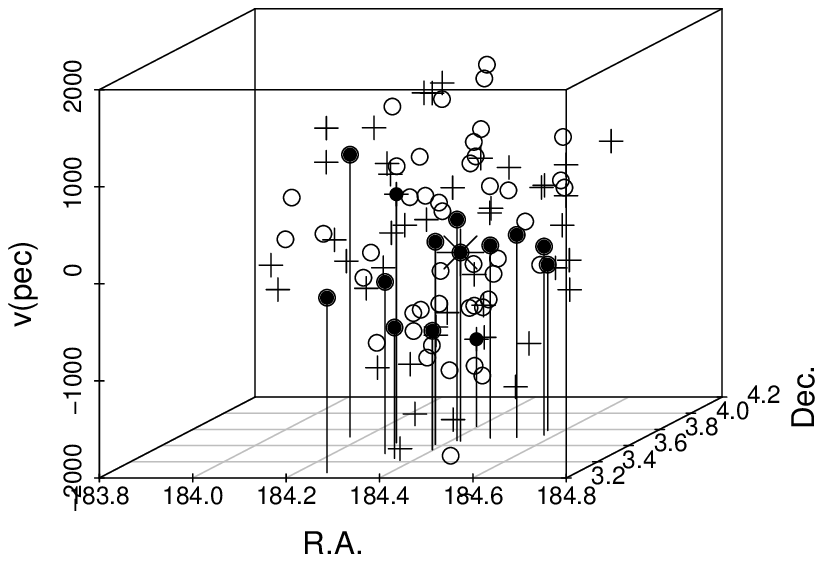}}}
\hspace*{2mm}\\
\caption{2D and 3D views of the cluster A1516 (SCl~111).
Symbols are the same as in Fig.~\ref{fig:gr225441}.
}
\label{fig:gr386211}
\end{figure}

The cluster A1516 is located in the high-density core of the supercluster SCl~111. 
The sky distribution of galaxies in this cluster shows very 
regular contours (Fig.~\ref{fig:gr386211}, left panel). According to 
{\it Mclust}, this cluster has one component, the mean uncertainty of the 
classification of galaxies  is $7.71\cdot10^{-4}$. 

\begin{figure}
\centering
{\resizebox{0.45\textwidth}{!}{\includegraphics*{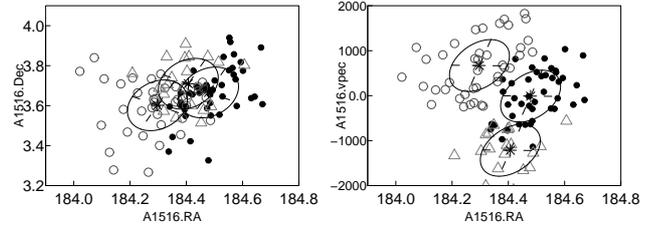}}}
\hspace*{2mm}\\
\caption{
Left panel shows R.A. vs. Dec., the right panel the  R.A. vs. the 
peculiar velocities of galaxies (in $km s^{-1}$)  in the cluster A1516. 
Different symbols correspond to different velocity components. 
}
\label{fig:gr386214}
\end{figure}

\begin{figure}
\centering
{\resizebox{0.30\textwidth}{!}{\includegraphics*{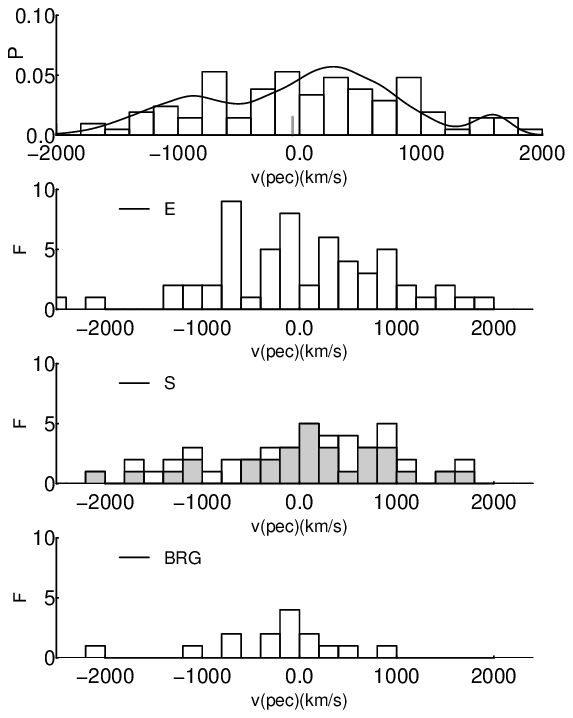}}}
\hspace*{2mm}\\
\caption{Distribution of the
peculiar velocities of galaxies in the cluster A1516. Top down: all galaxies,
elliptical galaxies, spiral galaxies, and BRGs.
The small dash in the upper panel indicates the peculiar velocity of the first 
ranked galaxy.
}
\label{fig:gr386212}
\end{figure}

The 3D distribution of galaxies shows that in the sky galaxies with negative peculiar 
velocities are mostly located in the central part of the cluster 
(Fig.~\ref{fig:gr386211}, right  panel, and Fig.~\ref{fig:gr386214}), where they form an 
extension with the outer parts delineated mostly by spiral galaxies. This 
component is seen also in Fig.~\ref{fig:gr386212}. According to the 
peculiar  velocities, the elliptical galaxies in this component are concentrated 
towards the centre of the cluster. There is another peak in the distribution of the 
peculiar velocities of elliptical galaxies closer to the cluster centre; a 
signature of a past merger? The distribution of the peculiar velocities of 
spiral galaxies is less peaky. The distribution of the peculiar velocities of red 
spiral galaxies shows a concentration at the centre, at low peculiar velocites. 
There is also a small subcluster of galaxies with high positive peculiar 
velocities. The gradient of the peculiar velocities of galaxies in A1516 
hints that this cluster may be rotating. With one exception, BRGs populate 
central parts of the cluster, and the first ranked galaxy of the cluster has a 
very low peculiar velocity, $v_{pec} = 58$ km/s. The $p$-value of the Shapiro-
Wilk test is 0.535, which confirms the 
Gaussian distribution of the peculiar velocities of galaxies in this cluster.

\begin{figure}
\centering
{\resizebox{0.35\textwidth}{!}{\includegraphics*{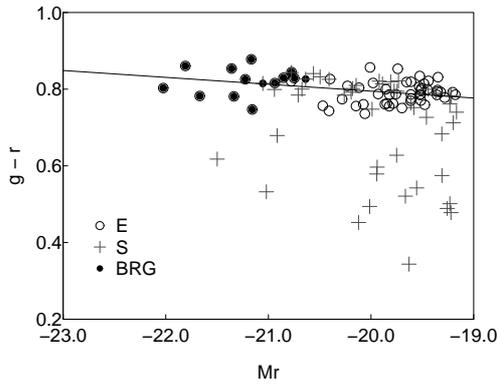}}}
\hspace*{2mm}\\
\caption{Colour-magnitude diagram of galaxies in the cluster A1516.
Empty circles denote elliptical galaxies, crosses  spiral galaxies,
filled circles  BRGs.
}
\label{fig:gr386213}
\end{figure}

The colour-magnitude diagram in Fig.~\ref{fig:gr386213} shows that the scatter 
of colours of red ellipticals and red spirals is almost equal. The fraction of 
red galaxies in this cluster is very high (85\%, Table~\ref{tab:grgal}), only 3 
of 15 BRGs in this cluster are classified as spirals.

\subsection{Clusters in the core of the supercluster SCl~126}

\subsubsection{The cluster A1650}

\begin{figure}
\centering
{\resizebox{0.22\textwidth}{!}{\includegraphics*{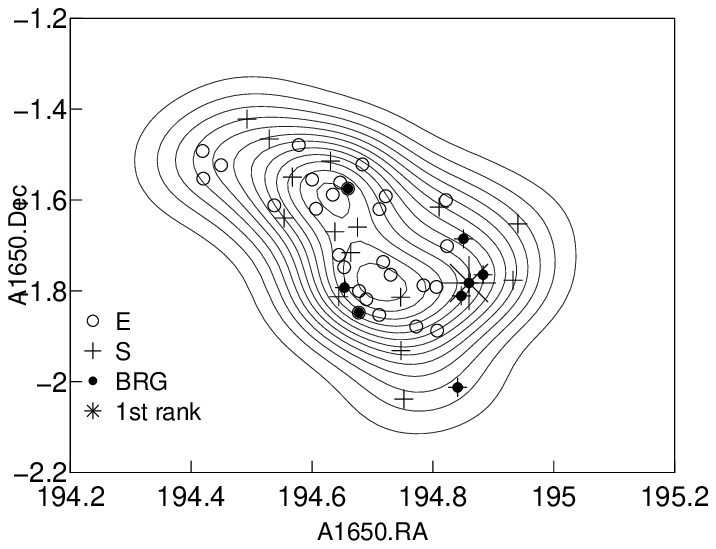}}}
{\resizebox{0.25\textwidth}{!}{\includegraphics*{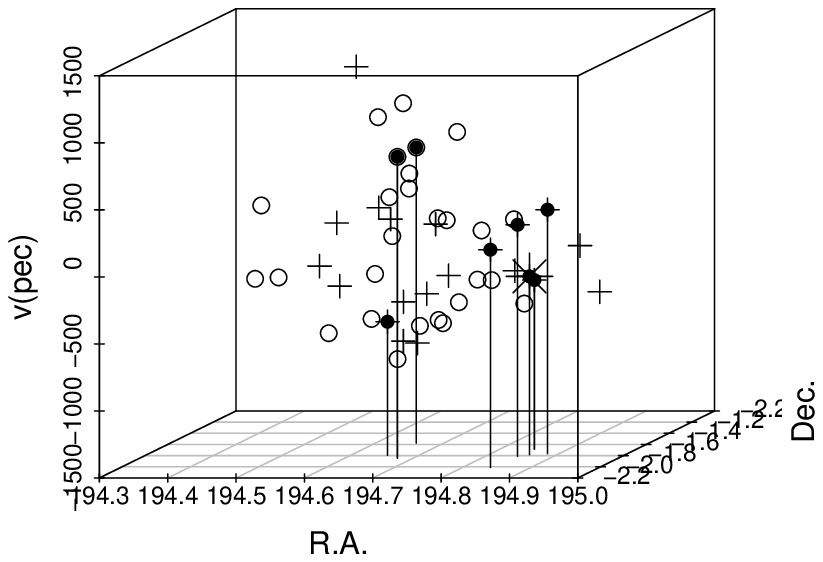}}}
\hspace*{2mm}\\
\caption{2D and 3D views of the cluster A1650.
Symbols are the same as in Fig.~\ref{fig:gr225441}.
}
\label{fig:gr446061}
\end{figure}

\begin{figure}
\centering
{\resizebox{0.30\textwidth}{!}{\includegraphics*{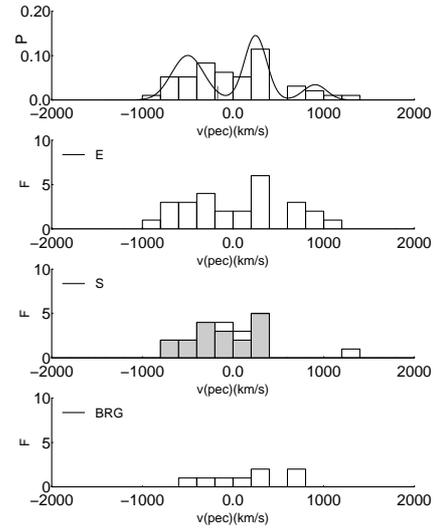}}}
\hspace*{2mm}\\
\caption{Distribution of the
peculiar velocities of galaxies  in the cluster A1650. Top down: all galaxies,
elliptical galaxies, spiral galaxies, and BRGs.
The small dash in the upper panel indicates the peculiar velocity of the first 
ranked galaxy.
}
\label{fig:gr446062}
\end{figure}

The first cluster we study in the core of the supercluster  SCl~126 is the 
cluster A1650, an X-ray cluster \citep{boe04,udo04}. 
 This cluster is the second-poorest cluster
in our sample with only 48 member galaxies in our catalogue.
The sky distribution of 
galaxies in this cluster (Fig.~\ref{fig:gr446061}, left panel) shows 
two concentrations. The 3D view and the velocity histograms of galaxies of 
different type (Fig.~\ref{fig:gr446061}, right panel, and Fig.~\ref{fig:gr446062}) 
show an almost separate component  of galaxies with positive peculiar velocities, 
all but one galaxies in this component are elliptical. In the sky distribution 
this component is located almost in the centre of the cluster, thus  
{\it Mclust}, using the data about all the galaxies, finds that this cluster 
consists of one component, with the mean uncertainty of the classification  
 0.008. Red spiral galaxies and BRGs (with one exception) 
populate the main component of the cluster. \citet{udo04} describes this cluster 
as a  compact X-ray source, possibly located at a cold spot in the CMB.

\begin{figure}
\centering
{\resizebox{0.22\textwidth}{!}{\includegraphics*{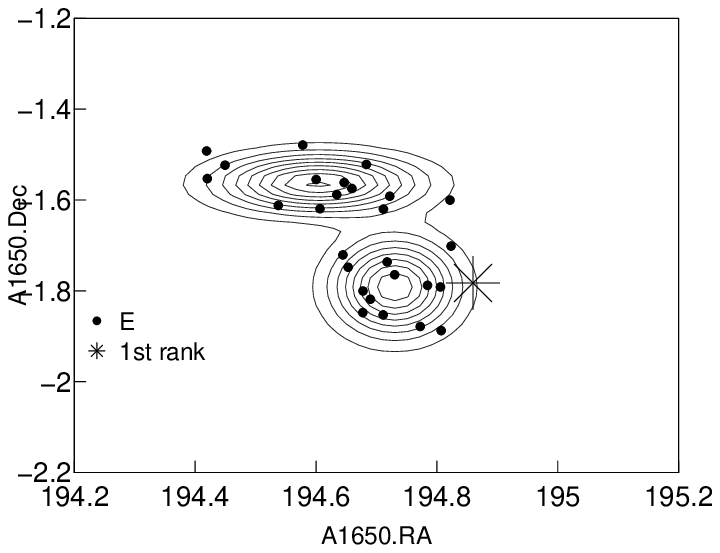}}}
{\resizebox{0.22\textwidth}{!}{\includegraphics*{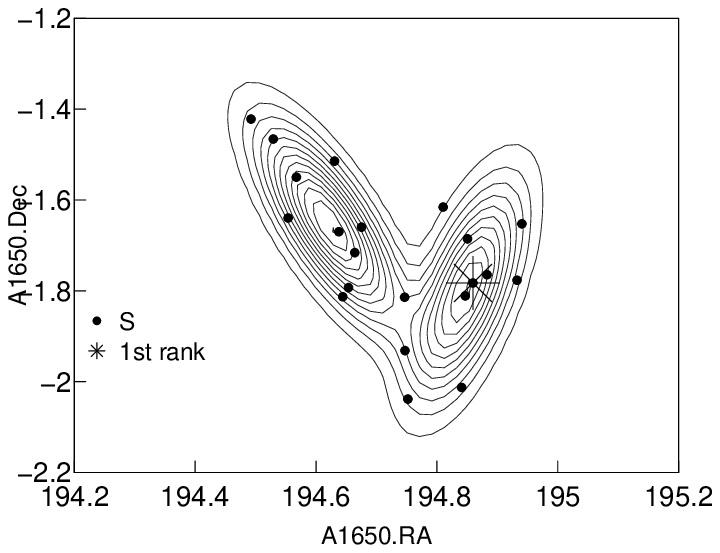}}}
\hspace*{2mm}\\
\caption{Distribution of elliptical
(left panel) and spiral (right panel) galaxies in the cluster A1650. 
Filled circles in left panel denote elliptical galaxies,
in right panel they show spiral galaxies. The star is the first ranked galaxy.
}
\label{fig:gr446064}
\end{figure}

However, the sky distribution of only elliptical or only spiral galaxies 
(Fig.~\ref{fig:gr446064}) clearly shows  two components 
in the distribution of  elliptical and spiral galaxies, in both cases in a 
different manner. Thus we approximated the distribution of the peculiar 
velocities of galaxies with three Gaussians. These are better seen in the 
distribution of the peculiar velocities of elliptical galaxies. The distribution 
of the peculiar velocities of spiral galaxies is different. There is one spiral 
galaxy only in one component, which has the highest peculiar velocity in this 
component. {\it Mclust} confirms the  components in the distribution of 
elliptical and spiral galaxies. The first ranked galaxy of this cluster is 
located at the edge of the elliptical galaxy component, 
but in the centre of one  spiral galaxy component; the peculiar 
velocity of the first ranked galaxy $v_{pec} = -189$ km/s.

\begin{figure}
\centering
{\resizebox{0.35\textwidth}{!}{\includegraphics*{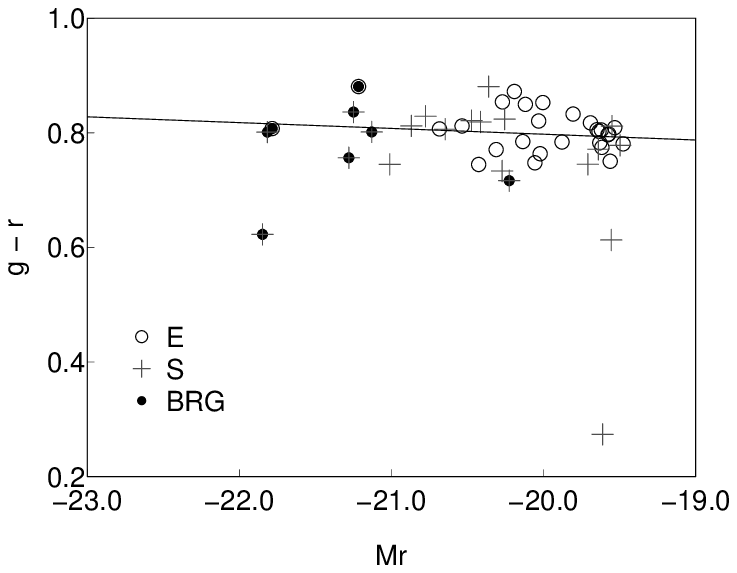}}}
\hspace*{2mm}\\
\caption{Colour-magnitude diagram of galaxies  in the cluster A1650.
Empty circles denote elliptical galaxies, crosses  spiral galaxies,
filled circles BRGs.
}
\label{fig:gr446063}
\end{figure}

The fraction of red galaxies in this cluster is
 very high, some spiral galaxies are also red. 
Owing to the small number of galaxies in this cluster the
high fraction of red galaxies is only approximate. 
The first ranked galaxy in 
this cluster is also a red spiral galaxy, as seen in the colour-magnitude diagram 
(Fig.~\ref{fig:gr446063}). In this diagram the rms scatter of colours of the red 
sequence is 0.036 for red ellipticals and 0.042 for red spirals.

Our analysis hints at a possibility
 that this cluster consists of two main components, 
some galaxies belong to a third subgroup with high positive peculiar velocities.
Because of the small number of galaxies in this cluster 
our results about substructures
should be taken as a suggestion only.

\subsubsection{The cluster A1658}

\begin{figure}
\centering
{\resizebox{0.22\textwidth}{!}{\includegraphics*{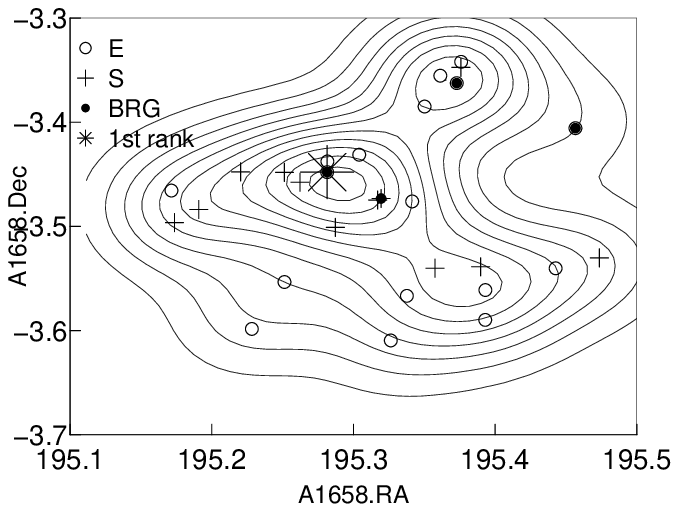}}}
{\resizebox{0.25\textwidth}{!}{\includegraphics*{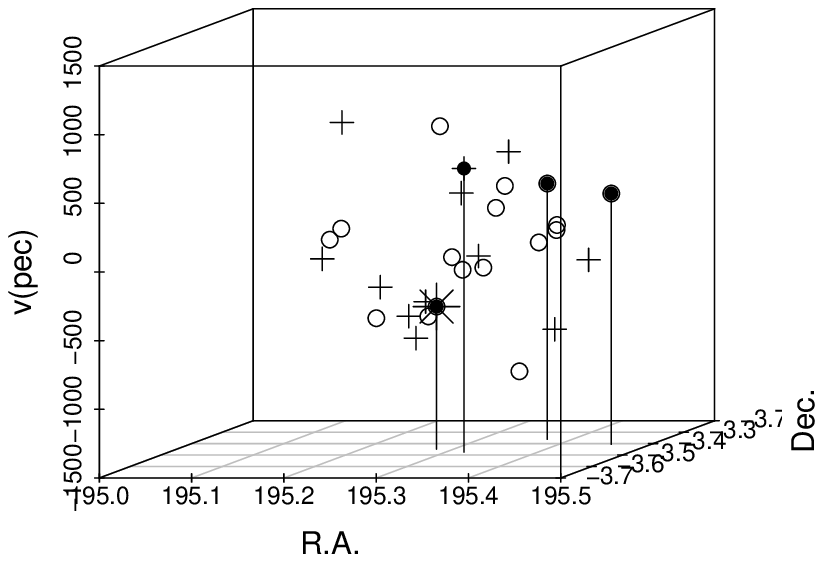}}}
\hspace*{2mm}\\
\caption{2D and 3D view of the cluster A1658.
Symbols are the same as in Fig.~\ref{fig:gr225441}.
}
\label{fig:gr449921}
\end{figure}

The next cluster we study in the core of the supercluster SCl~126 is the cluster 
A1658, the poorest cluster in our sample with only 29 member galaxies. The sky 
distribution of galaxies in this cluster (Fig.~\ref{fig:gr449921}, left panel) 
shows three irregularly located concentrations. According to 
{\it Mclust}, using the data about all the galaxies, this cluster has one component, 
the mean uncertainty of the classification of galaxies  is less than $10^{-5}$.

\begin{figure}
\centering
{\resizebox{0.30\textwidth}{!}{\includegraphics*{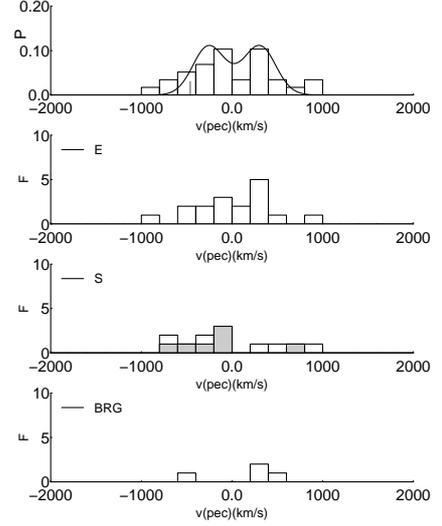}}}
\hspace*{2mm}\\
\caption{Distribution of the
peculiar velocities of galaxies  in the cluster A1658. Top down: all galaxies,
elliptical galaxies, spiral galaxies (grey -- red spirals), and BRGs.
The small dash in the upper panel indicates the peculiar velocity of the first 
ranked galaxy.
}
\label{fig:gr449922}
\end{figure}

The distribution of the peculiar velocities of galaxies in A1658 is shown in 
Fig.~\ref{fig:gr449922}. The distribution of the peculiar velocities of spiral 
galaxies shows two separate components, while the distribution of the 
peculiar velocities of elliptical galaxies has a central concentration, and two 
elliptical galaxies have high peculiar velocities. These distributions may be a 
signature of merging, which is more clearly seen in the distribution of the peculiar 
velocities of spiral galaxies. The $p$-value of the Shapiro-Wilk test is $p = 
0.937$, the highest among our cluster sample, confirms the Gaussianity of the
velocity distribution. The peculiar velocity of the first 
ranked galaxy in this cluster $v_{pec} = -463$ km/s; this galaxy is located in 
the richest component of this cluster. There are BRGs in both components, and most 
of the red spiral galaxies belong to one component.

\begin{figure}
\centering
{\resizebox{0.35\textwidth}{!}{\includegraphics*{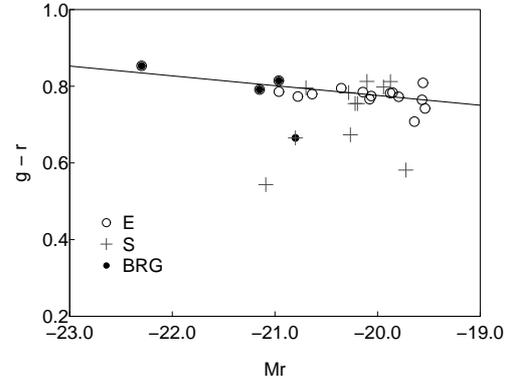}}}
\hspace*{2mm}\\
\caption{Colour-magnitude diagram of galaxies  in the cluster A1658.
Empty circles denote elliptical galaxies, crosses spiral galaxies,
filled circles  BRGs.
}
\label{fig:gr449923}
\end{figure}

 In the colour-magnitude diagram the scatter of colours of red galaxies in this 
cluster is very small, the fraction of red galaxies is high. Surprisingly, 
the scatter of colours of red spirals is even smaller than that of red 
ellipticals (with the rms values of 0.024 and 0.031, respectively). If the 
bimodal distribution of the peculiar velocities of galaxies in this cluster is an 
indication about a recent merger of two clusters into one, then the colour evolution
of galaxies was probably already finished in these clusters before merging.
 Again, as we also mentioned with regard to the cluster A1650,
the number of galaxies in this cluster is very small and 
our results concerning this cluster
should be taken as suggestion.

\subsubsection{The cluster A1663}

\begin{figure}
\centering
{\resizebox{0.22\textwidth}{!}{\includegraphics*{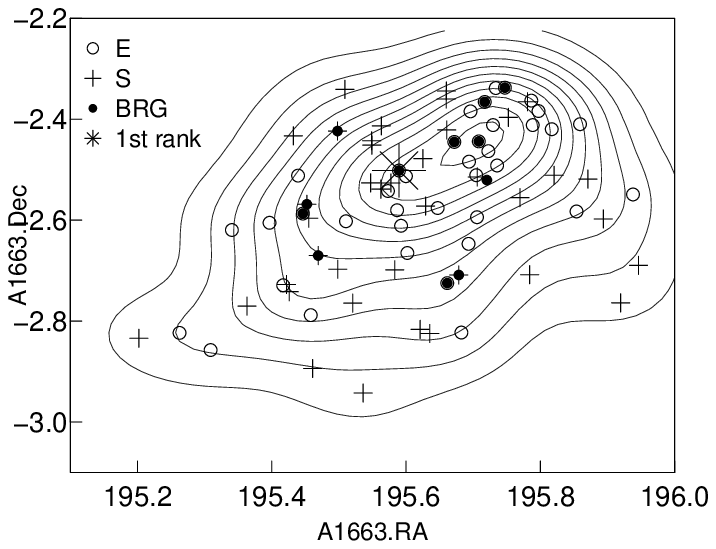}}}
{\resizebox{0.25\textwidth}{!}{\includegraphics*{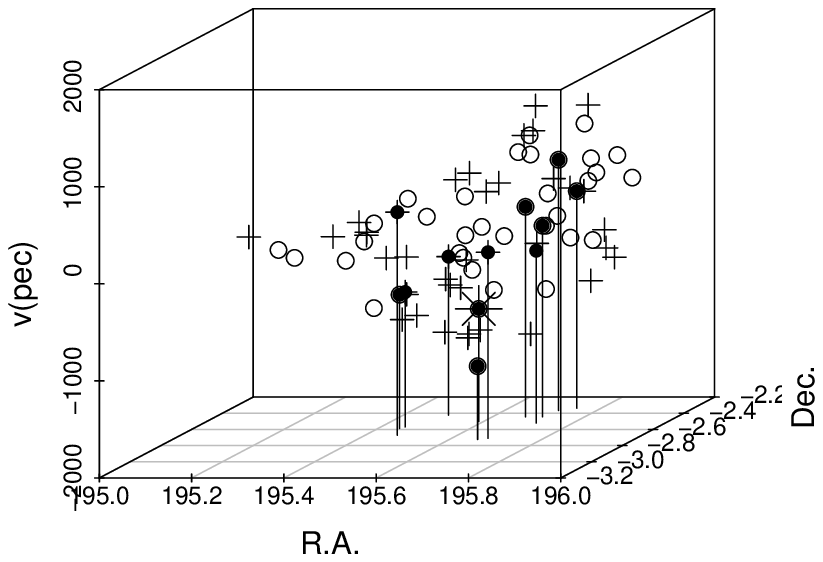}}}
\hspace*{2mm}\\
\caption{2D and 3D views of the cluster A1663.
Symbols are the same as in Fig.~\ref{fig:gr225441}.
}
\label{fig:gr450041}
\end{figure}

\begin{figure}
\centering
{\resizebox{0.45\textwidth}{!}{\includegraphics*{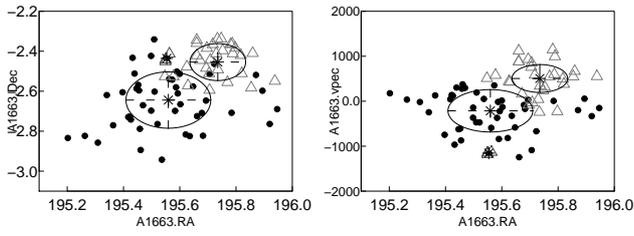}}}
\hspace*{2mm}\\
\caption{
Left panel shows  R.A. vs. Dec., the right panel the R.A. vs. 
peculiar velocities of galaxies (in $km s^{-1}$)  in the cluster A1663. 
Different symbols correspond to different velocity components. 
}
\label{fig:gr450044}
\end{figure}

\begin{figure}
\centering
{\resizebox{0.30\textwidth}{!}{\includegraphics*{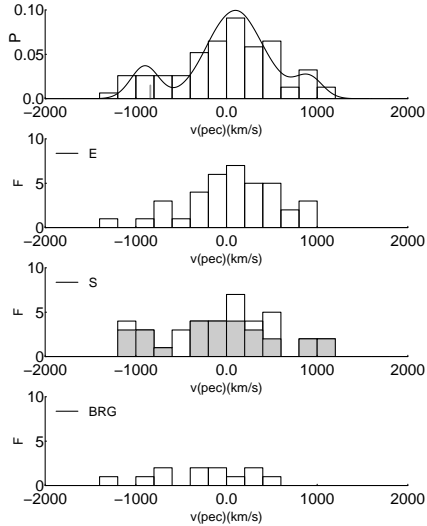}}}
\hspace*{2mm}\\
\caption{Distribution of the
peculiar velocities of galaxies  in the cluster A1663. Top down: all galaxies,
elliptical galaxies, spiral galaxies, and BRGs.
The small dash in the upper panel indicates the peculiar velocity of the first 
ranked galaxy.
}
\label{fig:gr450042}
\end{figure}

Figure~\ref{fig:gr450041} shows that the distribution of galaxies in the cluster 
A1663 has a concentration where mostly 
galaxies with positive peculiar velocities are located. These galaxies form a separate 
component in the velocity distribution (Fig.~\ref{fig:gr450042}).
According to {\it Mclust}, this cluster has two 
components with the mean uncertainty of the classification
$9.54\cdot10^{-3}$.

The distribution of the peculiar velocities of all galaxies in A1663 in 
Fig.~\ref{fig:gr450042} shows three components. A concentration of the peculiar 
velocities near the cluster centre is mostly due to elliptical galaxies.  The 
distribution of the peculiar velocities of spiral galaxies, especially that of 
red spirals, shows three subsystems, one of them corresponding to the 
component with high negative peculiar velocities. The $p$-value of the Shapiro-
Wilk test is $p = 0.37$, so it does not confirm the non-Gaussianity of the
velocity distribution. The BRGs in this cluster have peculiar velocities $v_{pec} 
< 500$ km/s, and in the sky distribution they are located at higher densities 
(according to the density contours in Fig.~\ref{fig:gr450041}, where the 
first ranked galaxy is also located. The first ranked galaxy in this cluster has a 
high negative peculiar velocity ($v_{pec} = -841$ km/s), suggesting that this 
cluster is not virialized yet. \citet{burgett04} suggest from the data from the 
2dFGRS that the velocity gradient in this cluster is an indication that the 
cluster may be rotating. 

\begin{figure}
\centering
{\resizebox{0.35\textwidth}{!}{\includegraphics*{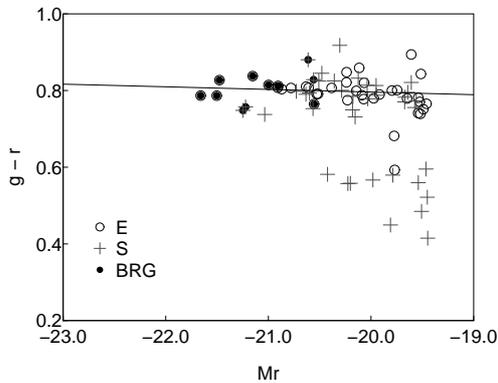}}}
\hspace*{2mm}\\
\caption{Colour-magnitude diagram of galaxies  in the cluster A1663.
Empty circles denote elliptical galaxies, crosses  spiral galaxies,
filled circles BRGs.
}
\label{fig:gr450043}
\end{figure}

The fraction of red galaxies in the cluster A1663 is high, 83\%, and 43\% of red 
galaxies are spirals. In the colour-magnitude diagram the red and blue galaxies 
in this cluster are clearly separated. The scatter of colours of red galaxies in 
the cluster A1663 is larger than in the cluster A1658, with the rms value 0.037, 
and the scatter of colours of red ellipticals is larger than that of red spirals 
(with the rms values of 0.032 and 0.042, correspondingly).

\subsubsection{The cluster A1750}

\begin{figure}
\centering
{\resizebox{0.22\textwidth}{!}{\includegraphics*{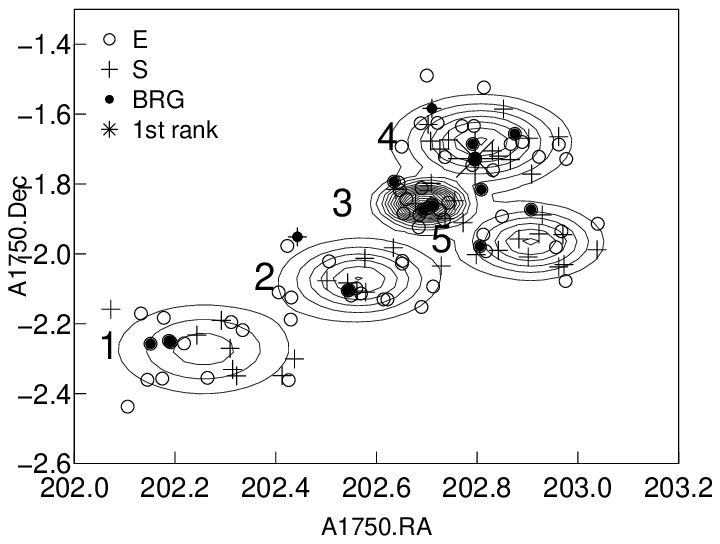}}}
{\resizebox{0.25\textwidth}{!}{\includegraphics*{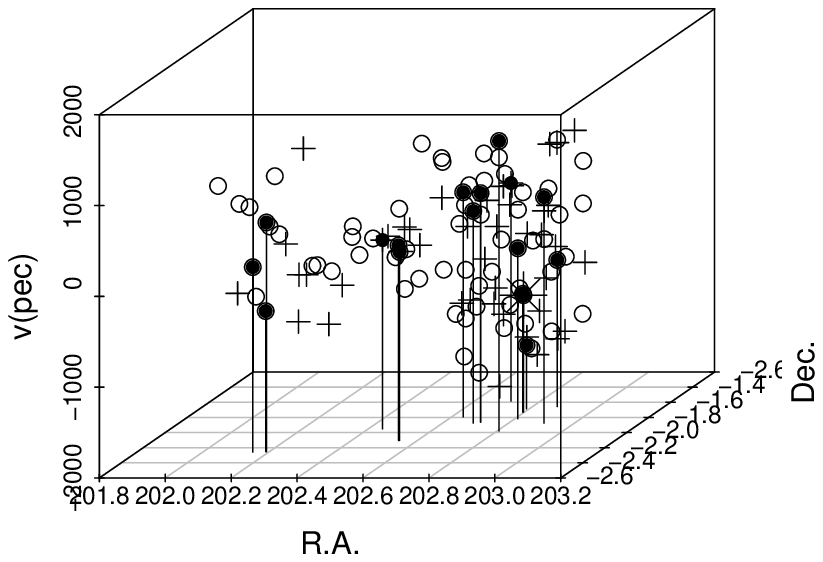}}}
\hspace*{2mm}\\
\caption{2D and 3D views of the cluster A1750.
Symbols are the same as in Fig.~\ref{fig:gr225441}.
}
\label{fig:gr491341}
\end{figure}

The next cluster, which is the richest cluster in our sample with 117 member galaxies 
in our group catalogue, is the well-known binary
merging Abell cluster A1750.

{\it Mclust} determined even five components in this cluster. In Fig.~\ref{fig:gr491341} 
(left panel) we mark them with numbers in an increasing order of the right 
ascencion. The mean uncertainty of the classification of galaxies in the cluster 
A1750  is $9.8\cdot10^{-3}$. 

\begin{figure}
\centering
{\resizebox{0.45\textwidth}{!}{\includegraphics*{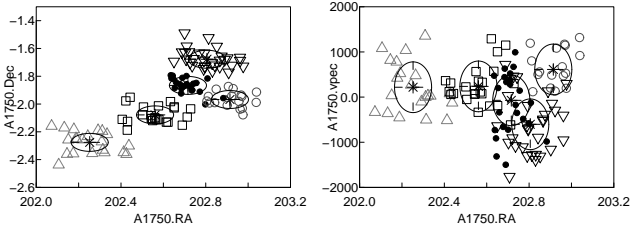}}}
\hspace*{2mm}\\
\caption{
Left panel shows R.A. vs. Dec., the right panel the R.A. vs. 
peculiar velocities of galaxies (in $km s^{-1}$)  in the cluster A1750. 
Different symbols correspond to different velocity components. 
}
\label{fig:gr491344}
\end{figure}

\begin{figure}
\centering
{\resizebox{0.30\textwidth}{!}{\includegraphics*{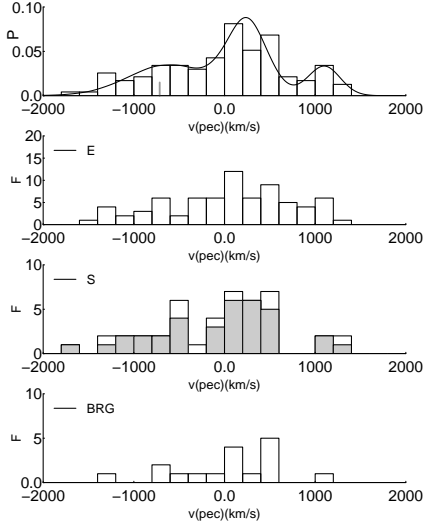}}}
\hspace*{2mm}\\
\caption{Distribution of the
peculiar velocities of galaxies  in the cluster A1750. Top down: all galaxies,
elliptical galaxies, spiral galaxies, and BRGs.
}
\label{fig:gr491342}
\end{figure}

The 3D distribution of galaxies in this cluster (Fig.~\ref{fig:gr491341}, right 
panel) shows two almost separate components. In one of them most galaxies have 
negative peculiar velocities with a mean value 
$v_{pec}= -600$ km/s. This component, marked as component 4 in the left panel of 
Fig.~\ref{fig:gr491341}, is well seen in the classification diagram generated with 
{\it Mclust}, Fig.~\ref{fig:gr491344}.  The first ranked galaxy of the cluster A1750 is 
also located in this component and has the peculiar velocity $v_{pec} = -724$ km/s.
In Fig.~\ref{fig:images} (left panel) we show an image of the first ranked galaxy
in this cluster. This galaxy, a BRG, is embedded in a common red halo
with its nearest companion galaxy which is the second in brightness in this cluster.
These galaxies may form a merging system. 
The histograms of the peculiar velocites (Fig.~\ref{fig:gr491342}) show that 
this component is better seen in the distribution of spiral galaxies than in the 
distribution of elliptical galaxies. Some BRGs are also located in  component 4.

In Fig.~\ref{fig:gr491341} the 2D density contours show that the sky density 
of galaxies in the cluster A1750 is the highest in component 3. 
The mean peculiar velocity of galaxies in this component is $v_{pec} = -72$ km/s.
Several BRGs are located here. 

In another component, component 5, galaxies have positive peculiar velocities,
$v_{pec} = 612$ km/s. There are two BRGs in this component, and some 
elliptical and spiral galaxies. 

In two components, 1 and 2, the mean peculiar velocity of galaxies is 
almost equal, $v_{pec} \approx 220$ km/s, and the 2D density of galaxies
decreases in these components as we move farther away from component 3 with a
high 2D galaxy density.  Figure~\ref{fig:gr491341} shows that also some BRGs are
located in these components, in the outer parts of the cluster.

The distribution of the peculiar velocites of red
spiral galaxies (Fig.~\ref{fig:gr491342}) follows that of all spirals
(of course, most spirals in this cluster are red).

The fraction of red galaxies in this cluster is very high (91\%), and the rms scatter of 
colours of red elliptical and red spiral galaxies is almost equal, about 0.04.
The colours of galaxies in different components are rather similar.
In this respect the cluster A1750 differs from the cluster A1205, where one component
consists mostly of red galaxies.

\begin{figure}
\centering
{\resizebox{0.35\textwidth}{!}{\includegraphics*{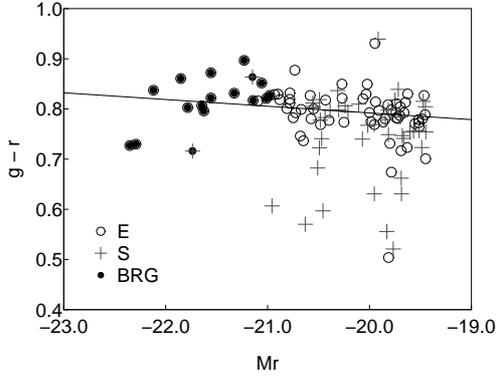}}}
\hspace*{2mm}\\
\caption{Colour-magnitude diagram of galaxies  in the cluster A1750.
Empty circles denote elliptical galaxies, crosses  spiral galaxies,
filled circles  BRGs.
}
\label{fig:gr491343}
\end{figure}

The structure of the cluster A1750 has been analysed in several other studies 
using optical and X-ray data \citep{don01,belsole04,burgett04,hwanglee09}. The 
comparison of substructures shows that our component 3 coincides with the 
component C by \citet{hwanglee09}. \citet{belsole04} finds an X-ray peak 
at this location and suggest that this component suffered a merger or 
interaction in the past 1--2 Gyrs. Our component 4 is the same as the one determined by 
\citet{hwanglee09} as component N, who suggest that the components 3 and 4 have
started interacting. Component 4 coincides with another X-ray peak 
determined by \citet{belsole04}. These are the same components denoted as 
Core and C in \citet{burgett04}, and as the NW and SW components in \citet{don01}. 
Our components 1 and 2 correspond to a sparse component S by 
\citet{hwanglee09}, \citet{belsole04} shows  an X-ray source in 
this direction. \citet{hwanglee09} suggest that there has been past 
interaction between the components 1, 2, and 3. The component E in 
\citet{hwanglee09} corresponds approximately to our component 5, and may have  
interacted in the past with both the components 3 and 4.
Hwang \& Lee did not find significant differences between the galaxy colours
in different components, similar to our study.

Thus in the cluster A1750 our analysis reveals a similar substructure to that
found in other studies.

\subsection{The cluster A1773 in the outskirts of the supercluster SCl~126}

\begin{figure}
\centering
{\resizebox{0.22\textwidth}{!}{\includegraphics*{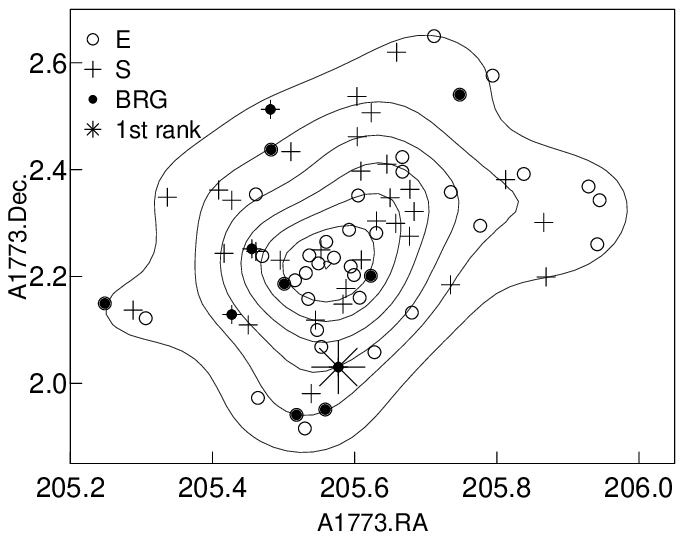}}}
{\resizebox{0.25\textwidth}{!}{\includegraphics*{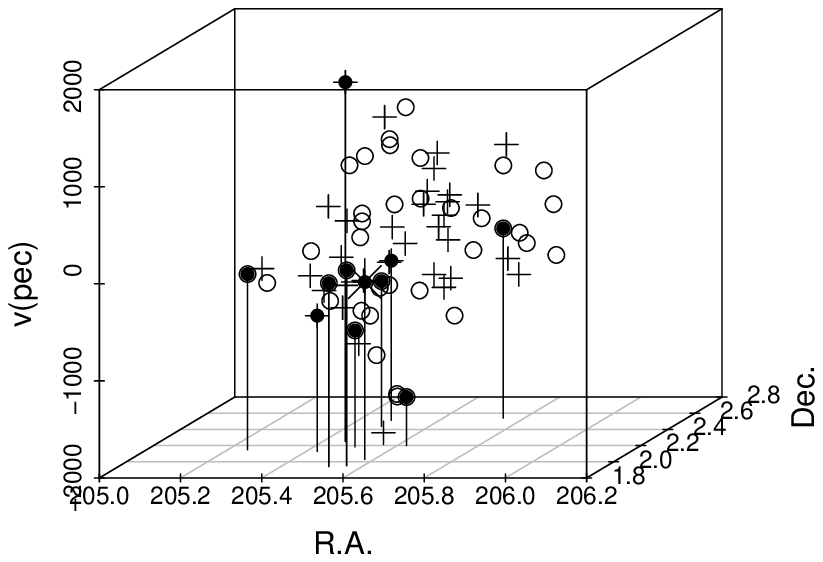}}}
\hspace*{2mm}\\
\caption{2D and 3D views of the cluster A1773.
Symbols are the same as in Fig.~\ref{fig:gr225441}.
}
\label{fig:gr509931}
\end{figure}

\begin{figure}
\centering
{\resizebox{0.30\textwidth}{!}{\includegraphics*{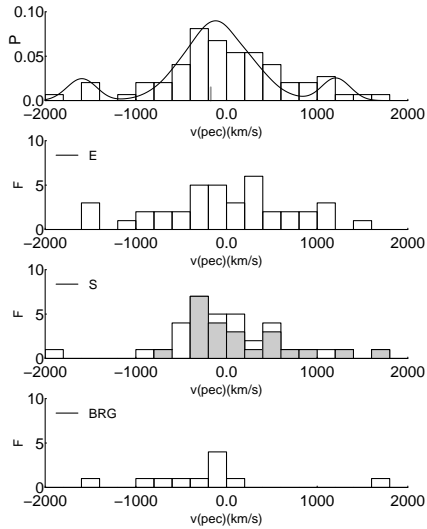}}}
\hspace*{2mm}\\
\caption{Distribution of the
peculiar velocities of galaxies  in the cluster A1773. Top down: all galaxies,
elliptical galaxies, spiral galaxies, and BRGs.
The small dash in the upper panel indicates the peculiar velocity of the first 
ranked galaxy.
}
\label{fig:gr509932}
\end{figure}

The next cluster in our sample is the X-ray cluster A1773, the member of the 
supercluster SCl~126, which is located in the outskirts of the supercluster. The sky 
distribution of galaxies in this cluster is irregular, especially in the outer 
regions (Fig.~\ref{fig:gr509931}, left panel). The 3D distribution of galaxies 
(Fig.~\ref{fig:gr509931}, right panel) shows  a component or tail 
of mostly elliptical galaxies and BRGs with  negative peculiar velocities. The 
histograms of the peculiar velocities  (Fig.~\ref{fig:gr509932}) confirm that 
the brightest galaxies in this cluster have mostly negative peculiar velocities. 
They are located in one part of the sky distribution (see 
Fig.~\ref{fig:gr509931}, left panel).The first ranked galaxy of the cluster is 
also located here, but has only a low peculiar velocity ($v_{pec} = -76$ 
km/s). However, this subsystem of galaxies does not form a completely separate 
component; according {\it Mclust}, this cluster has one component with the mean 
uncertainty of the classification 0.041. The distribution of the peculiar 
velocities of spiral galaxies in this cluster is smoother than that of 
elliptical galaxies. 

\begin{figure}
\centering
{\resizebox{0.35\textwidth}{!}{\includegraphics*{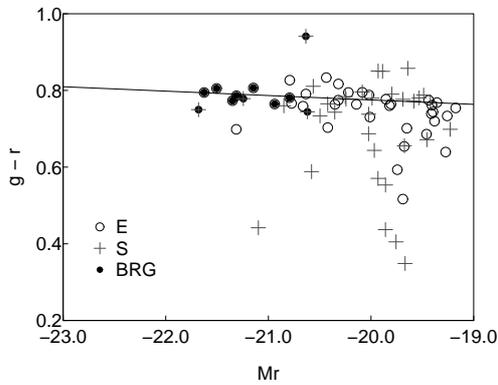}}}
\hspace*{2mm}\\
\caption{Colour-magnitude diagram of galaxies  in the cluster A1773.
Empty circles denote elliptical galaxies, crosses  spiral galaxies,
filled circles  BRGs.
}
\label{fig:gr509933}
\end{figure}

The fraction of red galaxies in the A1773 is 0.67 only. The scatter of colours 
of red ellipticals is smaller than that of red spirals (with the rms values of 
0.032 and 0.047, correspondingly, see Table~\ref{tab:grgal} and 
Fig.~\ref{fig:gr509933}). If galaxies with negative peculiar velocities have 
only recently joined this cluster  they have probably finished their colour 
evolution before joining the main cluster.

\subsection{The cluster A1809 in the outskirts of the supercluster SCl~126}

\begin{figure}
\centering
{\resizebox{0.22\textwidth}{!}{\includegraphics*{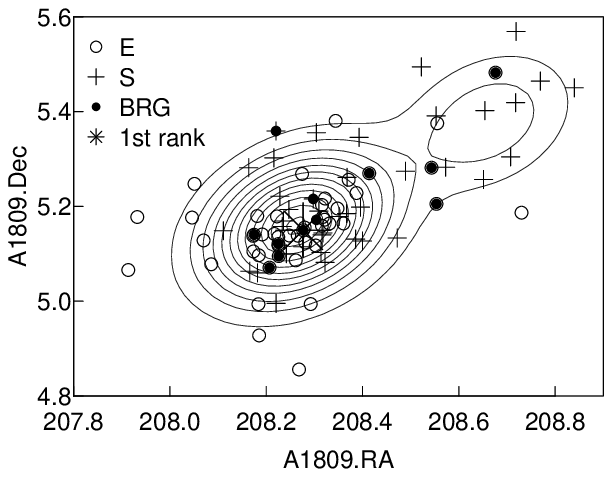}}}
{\resizebox{0.25\textwidth}{!}{\includegraphics*{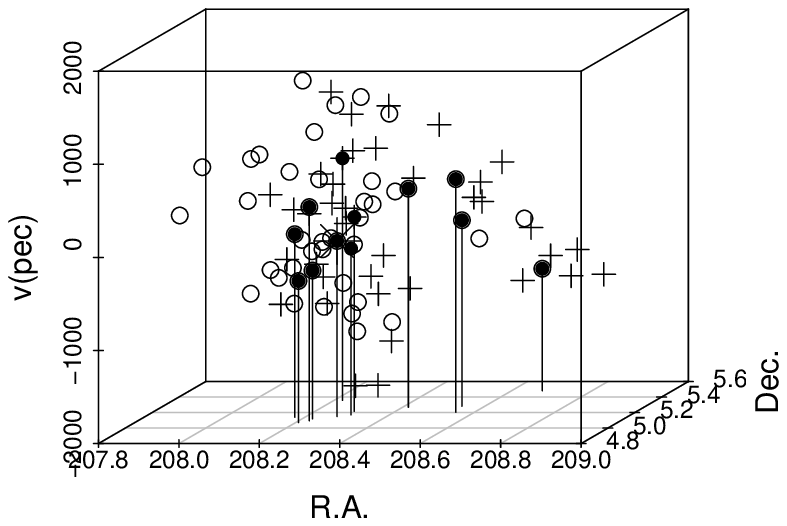}}}
\hspace*{2mm}\\
\caption{2D and 3D views of the cluster A1809.
Symbols are the same as in Fig.~\ref{fig:gr225441}.
}
\label{fig:A18091}
\end{figure}

\begin{figure}
\centering
{\resizebox{0.30\textwidth}{!}{\includegraphics*{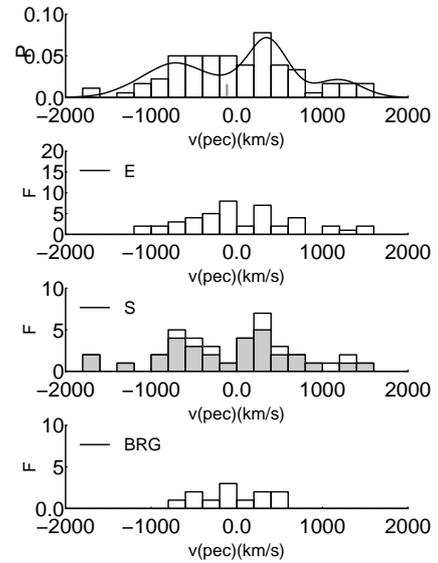}}}
\hspace*{2mm}\\
\caption{Distribution of the
peculiar velocities of galaxies  in the cluster A1809. Top down: all galaxies,
elliptical galaxies, spiral galaxies, and BRGs.
The small dash in the upper panel indicates the peculiar velocity of the first 
ranked galaxy.
}
\label{fig:A18092}
\end{figure}

The last cluster in our sample is the cluster A1809 in the outskirts of the 
supercluster SCl~126. The sky distribution of galaxies in this cluster 
(Fig.~\ref{fig:A18091}, left panel) shows that this cluster consists of two 
components in the sky. {\it Mclust} also confirms that this cluster consists of 
two components with the mean uncertainty of classification 
$1.3\cdot10^{-2}$. 
In the main component  of the cluster galaxies are concentrated in the 
center, and most BRGs of the cluster are also located here. The other component 
contains mostly loosely located spiral galaxies; there are also two elliptical 
galaxies and three BRGs, two of them located between this and the main component of 
the cluster.  
However, a word of caution is needed - in the poor component the number of
galaxies is small, 17, the galaxies here are loosely distributed; it is possible
that these galaxies cannot be considered as forming 
a reliable separate component. 
The 3D distribution of galaxies in the cluster A1809 
(Fig.~\ref{fig:A18091}, right panel) and the distribution of the peculiar 
velocities of galaxies (Fig.~\ref{fig:A18092}) show that in the main component 
of the cluster the galaxies have a bimodal distribution of the peculiar velocities, which is
best seen in the distribution for spiral galaxies. The velocity histograms show that 
some elliptical and spiral galaxies form a tail in the distribution of the 
peculiar velocities with high ($v_{pec} > 1000$ km/s) values of the peculiar 
velocities. The distribution of the peculiar velocities of red spiral galaxies 
follows those of all spirals. The velocity dispersion of the BRGs in this cluster 
is small, these galaxies are concentrated near the centre of the cluster. The 
distribution of the peculiar velocities of galaxies suggests that this cluster 
has formed by merging of two groups, and the third one (the poor component in 
the cluster A1809) is perhaps still infalling.

\begin{figure}
\centering
{\resizebox{0.35\textwidth}{!}{\includegraphics*{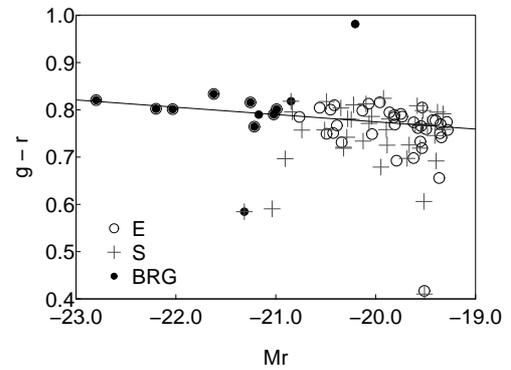}}}
\hspace*{2mm}\\
\caption{Colour-magnitude diagram of galaxies  in the cluster A1809.
Empty circles denote elliptical galaxies, crosses  spiral galaxies,
filled circles  BRGs.
}
\label{fig:A18093}
\end{figure}

The fraction of red galaxies in the cluster A1809 is 0.84, similar to that
in the clusters in the core of the supercluster. The scatter 
of colours of red ellipticals and red spirals (with the rms values of 0.029 and 
0.032, correspondingly, see Table~\ref{tab:grgal} and Fig.~\ref{fig:A18093})
is small. This suggests that the galaxies in the groups which formed the cluster A1809
possessed their final colours before merging.

The cluster A1809 has been studied for subclustering by \citet{1994AJ....107..857O,
oegerlehill2001}, who found no substructure here. The reason for the difference 
between their results and ours may be the difference in the data used -- in their
study the cluster includes a smaller number of galaxies, and the smaller component
is not well seen.  As we mentioned above, it is also possible that the
galaxies in the smaller component do not form a reliable 
separate component.

\section{Discussion}

\subsection{Dynamical state of rich clusters}

Our calculations with {\it Mclust} showed that among the ten clusters studied here five 
consist of more than one component, and the clusters A1205 and A1750 have the 
largest number of substructures. Even one-component clusters have several components in 
the distribution of the peculiar velocities of galaxies. In one cluster, A1205, 
one of the components is mostly populated by red galaxies, both ellipticals and 
spirals. In other clusters, the distribution of the peculiar velocities of 
galaxies shows that elliptical galaxies are  more concentrated towards the 
cluster centre or towards the centres of subclusters, than spiral galaxies. 
This is possibly a signature of recent mergers that affects the morphology of 
galaxies in cluster. In the cluster A1773 we see the opposite -- the velocity 
dispersion of spiral galaxies is lower than that of elliptical galaxies. This 
shows that the formation history of our clusters was different.

 A word of caution is needed concerning the poorest clusters in our 
sample, the clusters A1650 and A1658: owing to the small number of galaxies
in these clusters our results concerning them should be taken as a suggestion
only. Even clusters with smaller numbers of galaxies have been studied
for substructures \citep[see, for example,][]{1999A&A...343..733S,
2008A&A...487...33B};  however, \citet{2008A&A...487...33B} mention that 
clusters with about 30 member galaxies are too small for the analysis 
of the substructure \citep[see also the discussion about the substructure statistics 
in small samples in][]{2006A&A...456...23B}.

Clusters with multiple components belong to all three superclusters of the SGW. 
Based on our earlier studies of the supercluster SCl~126, we assumed that 
clusters in the high-density core of this supercluster are dynamically more 
evolved than clusters in other superclusters of the SGW, and close to be 
virialized. Now we see that this is not so, clusters in the high-density 
cores of superclusters also have substructure and thus are not yet virialized.

\citet{2007MNRAS.377.1785R} showed that in simulations group-size haloes located close to 
massive haloes have substructure. \citet{2004ogci.conf...19P} found that 
dynamically active clusters are more strongly clustered than the overall cluster 
population. Some studies of N-body models suggest that the fraction of haloes 
with substructure typically increases in high-density regions \citep[][and 
references therein]{2007MNRAS.377.1785R}. In our study all clusters lie in 
superclusters, but we did not detect clear differences in the properties of clusters 
in the supercluster cores and in the outskirts or in the poor supercluster SCl~91. 

We analysed the distribution of the peculiar velocities of galaxies in clusters. 
The number of galaxies in the clusters is not large, thus for the distribution of 
the peculiar velocities of galaxies we used the Shapiro-Wilk  test to check the 
agreement with normality. In Table~\ref{tab:grdyn} we give the $p$-values of this 
test. These values show that the highest probability that the distribution of 
the peculiar velocities of galaxies in a cluster is not normal, is recorded for the cluster 
A1066,  one-component cluster, in which the distribution of the peculiar 
velocities  of galaxies shows three components, a possible 
indication of merging subgroups. 

In Table~\ref{tab:grdyn} we give the values of kurtosis and skewness for the 
distributions of the peculiar velocities of galaxies in clusters
 \citep[see also][]{1999A&A...343..733S,
oegerlehill2001,2006A&A...449..461B, 
2007A&A...469..861B,2007ApJ...662..236H}. The positive 
values of kurtosis in Table~\ref{tab:grdyn} suggest that all distributions are 
peaked. The skewness shows that the distributions are asymmetrical, being skewed 
to  the right or left, depending on the cluster. However, the $p$-values are 
higher than 0.05 showing that deviations from the normal distribution are 
statistically not very significant. The $p$-value for kurtosis for the cluster A1205 is 
low, for skewness it is high. These tests compare different aspects of the 
distributions, thus the results may not be similar. We also see that the results 
of the tests alone are not always good indicators for the possible substructure in 
the distributions. We saw that in the clusters with several components in the sky
the distributions of the peculiar velocities of galaxies 
in different components sometimes (partly) coincide. Thus the low statistical significance 
for the deviations from normality 
is not surprising.
 
Another indicator of virialization of clusters is the peculiar velocity of 
the first ranked galaxy in cluster. In six clusters from our sample the peculiar 
velocities of the first ranked galaxies $v_{pec} > 180$ km/s. The relative peculiar 
velocities of the first ranked galaxies in our clusters, 
$|v_{pec}|/{\sigma}_{v}$ (Table~\ref{tab:grdyn}) can be divided into three classes: four
clusters have $|v_{pec}|/{\sigma}_{v} \leq 0.20$, two -- $|v_{pec}|/{\sigma}_{v} 
= 0.41$, and four clusters -- $|v_{pec}|/{\sigma}_{v} \geq 1.0$.  The low value 
of the relative peculiar velocity for the clusters A1205 and A1809
is misleading -- the 
presence of multiple components in these clusters is a clear evidence of the 
nonvirialized state. Two other clusters with relative peculiar velocities less 
than 0.20$\sigma_{v}$ are both one-component clusters. Both clusters with 
$v_{pec}$/$\sigma_{v} = 0.41$, although they are one-component systems, have several
components in the distribution of the peculiar velocities of galaxies. Four 
multicomponent clusters have the highest values of the relative peculiar 
velocities. 

High peculiar velocities of the first ranked galaxies in clusters have been 
found in several recent studies \citep{oegerlehill2001,coziol09}. Some of the 
clusters in our sample and in the sample of rich clusters in \citet{coziol09} 
coincide; for common clusters  our results agree well. 
 The high
peculiar velocities of the first ranked galaxies in clusters suggest
that those clusters have not been yet virialized after a recent merger
 \citep{malumuth92},
see also \citet{oegerlehill2001,coziol09}.  
Before merger, these galaxies were the first ranked galaxies of one group which 
merged to form a larger system. 
As evidence for that, we found that in several of 
our groups the first ranked galaxy is located near the centre of one component. 
In this respect the cluster A1205 is exceptional -- in this cluster the first 
ranked galaxy lies between different components.

All rich clusters we studied correspond to Abell clusters. An identification of 
clusters is usually complicated because of the differences between the data in the
catalogues. This is also the case here. We compared the data of our clusters with 
the data of Abell clusters from the catalogue by Andernach \& Tago (2009, 
private communication). This comparison showed that often several our groups can be 
associated with one Abell cluster. Our cluster, which we identified with the Abell 
cluster A1066, is the richest cluster among the four groups which can be 
associated with this cluster; among the other groups one has 35 member 
galaxies, others are pairs of galaxies. Other clusters from the catalogue by 
Andernach \& Tago (2009) can be associated with one rich cluster and one 
to four poor 
groups with up to five member galaxies. This also confirms the presence of 
substructure in rich clusters.

The cluster A1750 has been studied for substructure by other authors, too. We 
showed above that the components found in this cluster by us coincide 
well with those found in other studies. We found an especially good agreement 
between our results and those by \citet{hwanglee09}, who used used a $\Delta$ -
test \citep{dressler88} to determine the substructure 
in the cluster A1750. This shows that {\it Mclust}, probably applied here for the first time 
for the purpose to find substructure in galaxy clusters, gives results in 
agreement with other methods.

The good agreement with other studies also supports our choice of the parameters 
for the friend-of-friend 
algorithm for group definition and suggests that  in our catalogue groups  with 
substructure are real groups and not complexes of small groups artificially 
linked together by an unreasonable choice of linking parameters. This example 
supports our opinion that in the T10 the linking parameters were chosen 
reasonably.

An increasing number of studies have recently shown the presence of 
substructure in rich clusters 
\citep{1999A&A...343..733S,oegerlehill2001,burgett04, 
2006A&A...449..461B,2007A&A...469..861B,2007ApJ...662..236H,2010arXiv1007.3497A}
 and in poor clusters \citep{2008A&A...487...33B}; 
\citet{2009AN....330..998V} found a substructure also in poor groups. 
\citet{tov09} studied the properties  of poor groups from the SDSS survey and 
showed that many groups of galaxies presently are not in a dynamical equilibrium but 
at various stages of virialization. \citet{niemi} showed that a significant 
fraction of nearby groups of galaxies are not gravitationally-bound systems. 
This is important because  the masses of observed groups are often estimated 
assuming that groups are bound.

The presence of substructure in 
rich clusters, signs of possible mergers, infall, or rotation suggests that the rich 
clusters in our sample are not yet virialized.
The high frequency of these clusters tells us that mergers between groups and 
clusters are common -- galaxy groups continue to grow. 
In high-density regions groups and clusters of galaxies form early and
could be more evolved dynamically \citep{tempel09}, but here again the possibility
of mergers is high, thus groups and clusters in high-density regions
are still assembling.

\subsection{Galaxy populations in rich clusters}

\begin{figure}
\centering
{\resizebox{0.35\textwidth}{!}{\includegraphics*{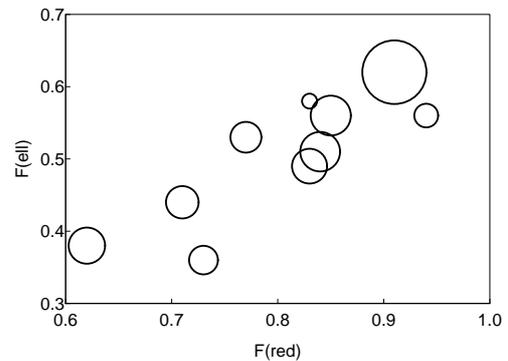}}}
\hspace*{2mm}\\
\caption{ Fractions of red galaxies and elliptical
galaxies for our clusters. Symbol sizes are proportional 
to the luminosity of clusters.
}
\label{fig:lre}
\end{figure}

The study of the galaxy content of the rich clusters showed in agreement with 
our earlier study \citep{e08} that the fraction of red galaxies in clusters from 
the core of the supercluster SCl~126 is very high, $F_{red} > 0.8$. However, 
this is as large as the fraction of red galaxies in the cluster A1809 from the 
outskirts of this supercluster, as well as in the cluster A1516 from the 
supercluster SCl~111.  
 We plot the fractions of red 
galaxies in clusters and the fractions of elliptical galaxies 
 in Fig.~\ref{fig:lre}. 
 This figure shows that the 
fraction of elliptical galaxies in clusters is proportional to the fraction of 
red galaxies. At the same time the luminosities of clusters vary strongly, i.e. the 
fraction of red galaxies in clusters from our sample does not correlate well with the
luminosity of clusters. In our earlier studies we found that 
the fraction of red galaxies is the highest in clusters in the cores of rich 
superclusters, both for low- and high-luminosity clusters
\citep{e07b}. 

We found that approximately  1/3 of red galaxies are spirals. Recent results 
of the Galaxy ZOO project  \citep{masters2009} showed that indeed a large 
fraction of red galaxies are spirals, and almost all massive galaxies are red 
independently of their morphological type. The colour-magnitude diagrams show 
that in some clusters (A1066, A1205, A1424) the scatter of the colours of red 
spirals is larger than the scatter of the colours of red ellipticals, suggesting that 
red ellipticals form a more homogeneous population than red spirals. 

The number of BRGs in our clusters varies from four in A1658, which is the poorest 
cluster in our sample, to 17 in the richest cluster, A1750. Some BRGs are 
spirals. BRGs can be found both in the central areas and in the outskirts
of clusters.
The result that BRGs lie in clusters is consistent with the strong small-
scale clustering of the LRGs \citep{zehavi04,eisenstein04,blake07} that has been 
interpreted in the framework of the halo occupation model, where more massive 
haloes host a larger number of LRGs \citep{2008arXiv0805.0002V,zheng08,tinker09,
2010ApJ...709..115W}.

The colours of galaxies in cluster environment may be  affected by many factors. 
Colours may reach their observed values during a  merging event when galaxies 
are affected by processes in the cluster environment 
\citep{2009ApJ...690.1292B}. For example, galaxies may be moved to the red 
sequence through the quenching  of star formation in blue galaxies in the 
cluster environment \citep{bower98,ruhland09,bamford2009,skelton2009}. As 
evidence for that we found that in some of the clusters in our study the scatter 
of colours of galaxies is large, probably due to merging of components that 
affects the colours of galaxies.

At the same time \citet{tinker09} mention in the analysis of the distribution of distant 
red galaxies in the halo model framework that distant red galaxies must 
have formed their stars before they become satellites in haloes. This is 
possible if these galaxies were members of poorer groups before the groups 
merged into a richer one. We also find that in some clusters the range of red 
colours of galaxies is very small although these clusters consisted of several 
(probably merging) components. Probably in these cases the galaxies 
have finished their colour evolution in small groups before merging.

\subsection{Merger analysis of dark matter haloes from simulation}

\begin{table*}[ht]
\caption{
The properties of the five most massive dark matter haloes}
 \label{tab2}
 \centering
\begin{tabular}{rrrrrrrrr} 
\hline 
(1)&(2)&(3)&(4)& (5)&(6)&(7)& (8)&(9)\\      
\hline 
 $H_{ID}$ & $N_{SH}$ & $M_{MH}$ & $MH_z$ & $R_{vir}$ & $mean_{SH(z)}$ &
 $\Sigma N_{merg}$ & $N_{major}$ &  $N_{clust}$  \\[6pt]

  & & [$10^{14} h^{-1} \mathrm{M}_{\odot}$] & & [$h^{-1}$ Mpc]&  &
 $z<0.25$ & $z<0.5$ $f>0.1$ & \\[6pt]

\hline
 0 & 46 & 2.09 & 0.778 & 1.24 & 3.39 (2.02) & 42  & 0 & 1 \\
 1 & 64 & 1.75 & 0.271 & 1.17 & 3.68 (2.13) & 60  & 3 & 2 \\
 2 & 35 & 1.19 & 0.695 & 1.03 & 2.93 (1.64) & 49  & 0 & 1 \\
 3 & 46 & 1.18 & 0.271 & 1.03 & 2.80 (1.76) & 56  & 1 & 1 \\
 4 & 29 & 1.07 & 0.957 & 0.994 & 2.95 (2.02) & 34 & 0 & 1 \\

\hline
\label{tab:halodata}                        
\end{tabular}\\
\small\rm\noindent Columns in the Table are as follows:
Col. 1: halo ID index,
Col. 2: number of subhaloes (SH) at $z=0$.
Col. 3: mass of the main halo at $z=0$ (in units of $10^{14} h^{-1} \mathrm{M}_{\odot}$),
Col. 4: the formation time, defined as the redshift when the halo
has half of its mass at the present epoch, 
Col. 5: size of the halo (virial radius, in units of $h^{-1}$ Mpc), 
Col. 6: mean value and standard deviation of the formation time of subhaloes,
Col. 7: total number of recent ($z<0.25$) merger events $\Sigma N_{merg}$,
Col. 8: number of major mergers at $z<0.5$,
Col. 9: number of
components in a halo at the present epoch 
found by {\it Mclust} as for observed clusters.
\end{table*}

\subsubsection {Description of simulations}

To illustrate the formation of present-day haloes via multiple mergers during 
their evolution in a cosmological simulations we present in this subsection 
merger histories of dark matter haloes in an N-body simulation. 

We ran a $\Lambda$CDM simulation to study the merging history 
of dark matter haloes in a cosmological simulation.
We used the GADGET-2 code \citep{2001NewA....6...79S,2005MNRAS.364.1105S}.
In the simulation the dark matter haloes were 
identified with an algorithm called AHF \citep[Amiga
Halo Finder][]{2009ApJS..182..608K}, which is based on the adaptive 
grid structure of the simulation code in AMIGA (Adaptive Mesh
Investigations of Galaxy Assembly). All haloes consist of the
main halo -- the most massive halo in the group, and subhaloes located
 within the virial radius sphere of the main halo.

For the simulation we adopted a cosmological $\Lambda$CDM model with 
($\Omega_m+\Omega_{\Lambda}+\Omega_b=1$) and $h = 0.71$, 
the dark matter density $\Omega_{dm}=0.198$, the baryonic density 
$\Omega_b=0.042$, the vacuum energy density $\Omega_{\Lambda}=0.76$ and the rms 
mass density fluctuation parameter $\sigma_8=0.77$. The simulation has  $512^3$ 
dark matter particles, the  volume of the simulation is 80$h^{-1}$ Mpc~$^3$. In 
this simulation the masses of the largest haloes are of the order of $10^{14} 
h^{-1} \mathrm{M}_{\odot}$, the mass of an individual DM particle is 
2.54$\times 10^8 h^{-1}\mathrm{M}_{\odot}$. Our requirement for halo 
identification is that it has at least 100 particles, and therefore the minimum 
mass for haloes is 2.5$\times 10^{10} h^{-1}\mathrm{M}_{\odot}$. 
In the simulation, the main haloes with embedded subhaloes 
correspond to the observed groups and clusters of galaxies.
Studies of substructure in observed groups and clusters 
of galaxies and in dark matter haloes in a similar manner provide an important 
probe for the formation of observed galaxy systems.

\begin{figure}
\centering
{\resizebox{0.35\textwidth}{!}{\includegraphics*{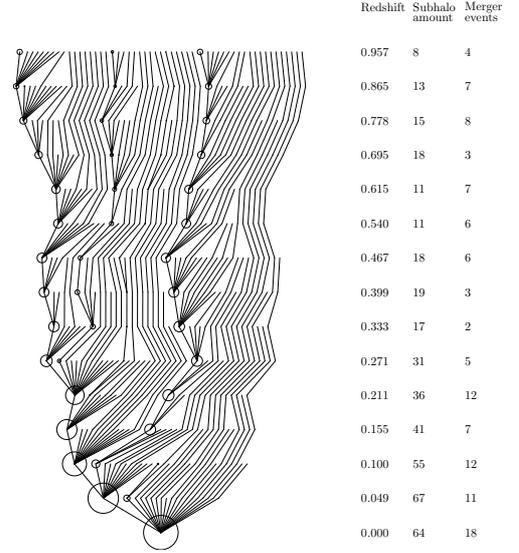}}}
\hspace*{2mm}\\
\caption{Complete merger tree for a halo with ID = 1. Major merger events are indicated
by circles with the size proportional to the mass of the merged halo.
}
\label{fig:mer1}
\end{figure}

\begin{figure}
\centering
{\resizebox{0.35\textwidth}{!}{\includegraphics*{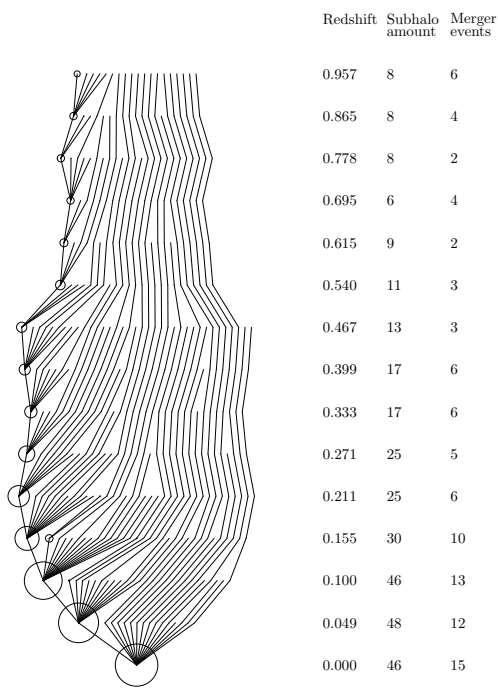}}}
\hspace*{2mm}\\
\caption{Complete merger tree for a halo with ID = 3. 
Merger events as indicated in Fig.~\ref{fig:mer1}.
}
\label{fig:mer3}
\end{figure}

\begin{figure}
\centering
{\resizebox{0.22\textwidth}{!}{\includegraphics*{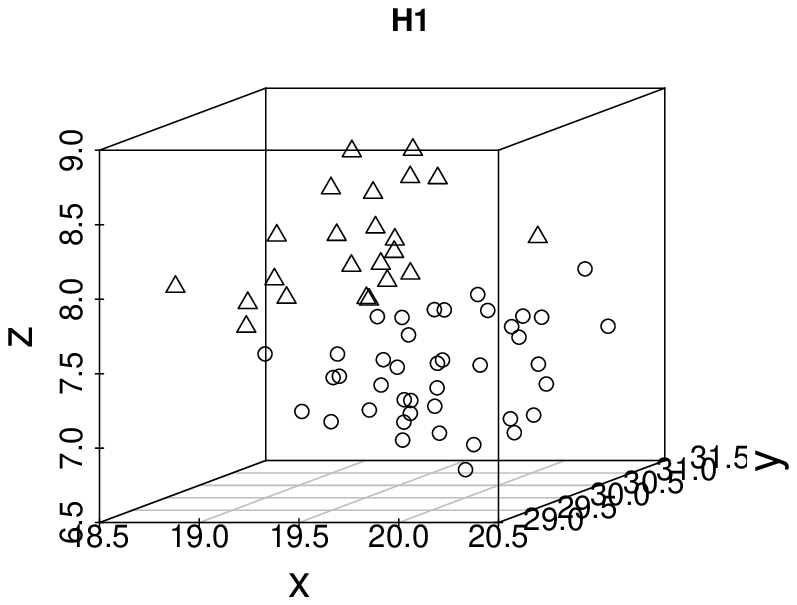}}}
{\resizebox{0.22\textwidth}{!}{\includegraphics*{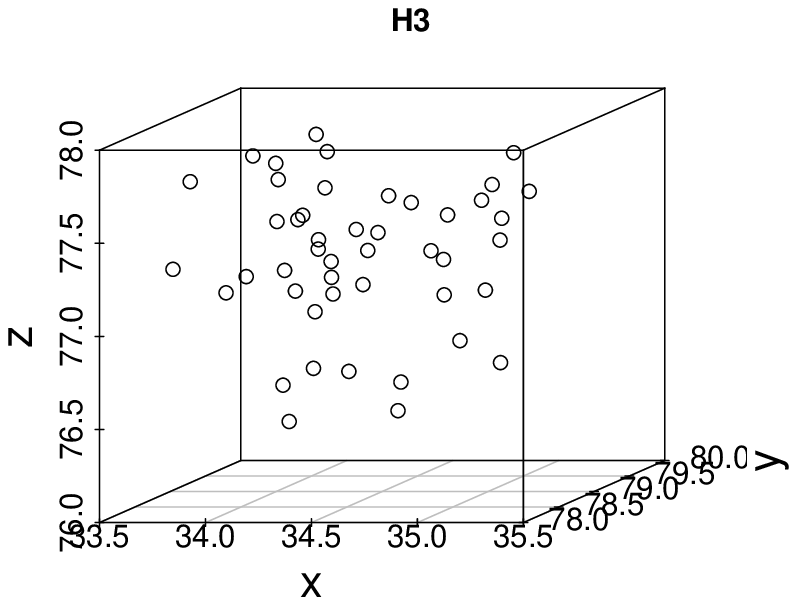}}}
\hspace*{2mm}\\
\caption{{\it Mclust} results for haloes 1 and 3: 3D view of the components.
}
\label{fig:halo13mclust}
\end{figure}

\subsubsection {Halo mergers}

The hierachical formation theory predicts that small haloes form first and large 
haloes are later assembled via mergers. To study the mergers between haloes 
during their formation, we calculated complete merger histories for the five 
richest (most massive) main haloes in the simulation.  We focused on the late-time 
evolution, because it is more likely that the events occuring during this epoch 
may still be seen in the structure of the haloes. 

In Table~\ref{tab:halodata} we show the results of our analysis. We calculated 
the formation times for the main haloes, defined as the redshift when halo has half 
of its mass at the present epoch, and the mean value of the formation time for 
all the subhaloes with the standard deviation.  We also show in 
Table~\ref{tab:halodata} 
the total 
number of merger events for main haloes for $z<0.25$ and the number of major 
mergers where the mass of the merging halo is at least 0.1 of the main halo mass.

To illustrate the complexity of halo histories, we show two examples of merger 
trees that we used for calculating the properties of mergers. In 
Fig.~\ref{fig:mer1} we show the merger tree for the halo ID~1, which had three 
large recent merger events (at redshifts $z$ = 0.0, 0.049, and 0.211). This halo 
has recently accreted its mass and its formation time is $z=0.271$. This can be 
compared with the merger history for the halo ID~3, which has the same formation 
time and almost the same amount of recent mergers, but in this halo only one 
large merger event at redshift $z = 0.1$ is clearly visible. Thus the merger 
tree of halo ID~3 is quite different from that of halo ID~1 
(Fig.~\ref{fig:mer3}). Interestingly, the mean value of the formation time for 
subhaloes in the halo ID~3 is lower than that of the halo ID~1. 
The number of components found by {\it Mclust} is 2 
for the halo ID~1 and 1 for the halo ID~3 (Table~\ref{tab:halodata} and 
Fig.~\ref{fig:halo13mclust}). Qualitatively one can say that more active 
late-time merging is visible in the clustering properties of haloes at $z=0$.    

A similar analysis was carried out for the five most massive simulated
haloes. Our results show 
that a late-time formation of the main haloes and a number of recent major 
mergers can cause a late-time subgrouping of haloes. A large scatter in the 
formation times for subhaloes hints towards late time merging (the standard 
deviation in the Col. 6). In general, we notice that the number of recent 
large merger events is the best indicator for additional clustering and that it is 
strongly correlated with the {\it Mclust} findings. To better understand the 
merging histories of dark matter haloes a study of a larger sample of haloes is 
needed.

\section{Conclusions}
We studied the properties of rich clusters in different superclusters
in  the  SGW. Our conclusions are as follows.

\begin{itemize}
\item[1)]
We showed with {\it Mclust}, a publicly available package in the $R$
statistical environment, that the richest clusters in all
superclusters of the SGW have substructure. 
\item[2)] 
There are components in the distribution of the peculiar velocities of 
galaxies in clusters both  with multiple components
and in one-component clusters -- evidence of merging components.
\item[3)] 
The peculiar velocities of the first ranked galaxies in clusters with
multiple components are high. 
\item[4)]
Most rich clusters in the supercluster SCl~126 and 
the richest cluster in the supercluster SCl~111
have a very high fraction of red
galaxies ($F_r > 80$ \%). About 1/3 of red galaxies in clusters are spirals.
In several clusters the scatter of colours of red elliptical galaxies in the
colour-magnitude diagram is smaller than the scatter of colours of red 
spirals, suggesting that red elliptical galaxies form a more homogeneous
population than red spiral galaxies.
The presence of colour-related substructure in these clusters suggests 
that the colour evolution of
galaxies in these clusters has already finished  before merging into
clusters.
\item[5)]
Our clusters host a large number of BRGs. Some BRGs are spirals. 
They can be found both in the central areas and in the outskirts of clusters.
\item[6)]
We calculated the complete merger trees for the richest dark matter 
haloes in an N-body simulation and showed that 
the late-time formation of the main haloes and the
amount of recent major mergers can cause the late-time subgrouping of
haloes.
\end{itemize}

Our results  about the presence of substructure in 
rich clusters, signs of possible mergers, infall or rotation  suggest that rich 
clusters in different superclusters of the Sloan Great Wall,
including the cores of superclusters, are not yet virialized. 
The richest clusters in the SGW are still assembling.

As a next step in the study of substructures in groups and clusters, we plan to study 
a larger sample of groups and clusters, including poor groups of galaxies and 
low-density global environments.

\section*{Acknowledgments}

We are pleased to thank the SDSS Team for the publicly available data
releases.  
Funding for the Sloan Digital Sky Survey (SDSS) and SDSS-II has been
  provided by the Alfred P. Sloan Foundation, the Participating Institutions,
  the National Science Foundation, the U.S.  Department of Energy, the
  National Aeronautics and Space Administration, the Japanese Monbukagakusho,
  and the Max Planck Society, and the Higher Education Funding Council for
  England.  The SDSS Web site is \texttt{http://www.sdss.org/}.
  The SDSS is managed by the Astrophysical Research Consortium (ARC) for the
  Participating Institutions.  The Participating Institutions are the American
  Museum of Natural History, Astrophysical Institute Potsdam, University of
  Basel, University of Cambridge, Case Western Reserve University, The
  University of Chicago, Drexel University, Fermilab, the Institute for
  Advanced Study, the Japan Participation Group, The Johns Hopkins University,
  the Joint Institute for Nuclear Astrophysics, the Kavli Institute for
  Particle Astrophysics and Cosmology, the Korean Scientist Group, the Chinese
  Academy of Sciences (LAMOST), Los Alamos National Laboratory, the
  Max-Planck-Institute for Astronomy (MPIA), the Max-Planck-Institute for
  Astrophysics (MPA), New Mexico State University, Ohio State University,
  University of Pittsburgh, University of Portsmouth, Princeton University,
  the United States Naval Observatory, and the University of Washington.

We thank our referee for detailed and constructive
comments and suggestions which helped to improve the paper.  
The present study was supported by the Estonian Science Foundation
grants No. 8005, 7146 and 7765, and by the Estonian Ministry for Education and
Science research project SF0060067s08. This work has also been supported by
the University of Valencia through a visiting professorship for Enn Saar and
by the Spanish MEC project AYA2006-14056 , ``PAU'' (CSD2007-00060), including
FEDER contributions, and the Generalitat Valenciana project of excellence 
PROMETEO/2009/064.  J.E.  thanks
Astrophysikalisches Institut Potsdam (using DFG-grant 436 EST 17/4/06), 
where part of this study was performed.  
P.H. was supported by the Societas Scientiarum Fennica, and Jenny 
  and the Antti Wihuri foundation. P.N. was supported by the Academy of Finland. 
The cosmological simulations were run at the CSC - Scientific Computing 
center in Finland.

\bibliographystyle{aa}

\end{document}